\documentclass{article}
\usepackage[english]{babel}
\usepackage[cp1251]{inputenc}
\begin{document}
\begin{center}
\textbf{\Large{Applications of an operator $H\left( {\alpha ,\beta
} \right)$ to the Lauricella multivariable hypergeometric
functions}}\\
\medskip
\textbf{A. Hasanov}\\
\medskip
Institute of Mathematics and Information Technologies, Uzbek
Academy of Sciences,\\ 29, F.
Hodjaev street, Tashkent 100125, Uzbekistan\\
E-mail: anvarhasanov@yahoo.com
\end{center}
\medskip
\begin{abstract}

By making use of some techniques based upon certain inverse new
pairs of symbolic operators, the author investigate several
decomposition formulas associated with Lauricella's hypergeometric
functions $F_A^{\left( r \right)}, F_B^{\left( r \right)},
F_C^{\left( r \right)}$ and $F_D^{\left( r \right)}$ in $r$
variables. In the three-variable case some of these operational
representations are constructed and applied in order to derive the
corresponding decomposition formulas when $r = 3$ . With the help
of these new inverse pairs of symbolic operators, a total 20
decomposition formulas and integral representations are found.\\
\textbf{MSC: primary 33C65.}\\
\textbf{Key Words and Phrases.} Decomposition formulas; Lauricella
hypergeometric functions; Multiple hypergeometric functions;
Generalized hypergeometric functions; Inverse pairs of symbolic
operators;  Integral representations.

\end{abstract}

\section{Introduction and definitions}

A great interest in the theory of hypergeometric functions (that
is, hypergeometric functions of several variables) is motivated
essentially by the fact that the solutions of many applied
problems involving (for example) partial differential equations
are obtainable with the help of such hypergeometric function (see,
for details, [21, p. 47]; see also the recent works [7-9, 16, 17]
and the references cited therein). For instance, the energy
absorbed by some nonferromagnetic conductor sphere included in an
internal magnetic field can be calculated with the help of such
functions [12, 15]. Hypergeometric functions of several variables
are used in physical and quantum chemical applications as well
[13, 19 and 20]. Especially, many problems in gas dynamics lead to
solutions of degenerate second-order partial differential
equations, which are then solvable in terms of multiple
hypergeometric functions. Among examples, we can cite the problem
of adiabatic flat-parallel gas flow without whirlwind, the flow
problem of supersonic current from vessel with flat walls, and a
number of other problems connected with gas flow [2, 6]. \\
We note that Riemann's functions and fundamental solutions of the
degenerate second-order partial differential equations are
expressible by means of hypergeometric functions of several
variables [7-9]. In investigation of the boundary value problems
for these partial differential equations, we need decompositions
for hypergeometric functions of several variables in terms of
simpler hypergeometric functions of the Gauss and Lauricella
types. \\
Multiple hypergeometric functions (that is, hypergeometric
functions in several variables) occur naturally in a wide variety
of problems. In particular, the Lauricella functions $F_A^{\left(
r \right)}$, $F_B^{\left( r \right)}$, $F_C^{\left( r \right)}$
and $F_D^{\left( r \right)}$ in $r$ $(r \in \backslash \left\{ 1
\right\};\,: = \left\{ {1,2,...} \right\})$ variables, defined by
( [1, p. 114])
$$
\begin{array}{l}
F_A^{\left( r \right)} \left( {\alpha ;\beta _1 ,...,\beta _r ;\gamma _1 ,...,
\gamma _r ;x_1 ,...,x_r } \right) = \sum\limits_{m_1 ,...,m_r  = 0}^\infty  {}
\displaystyle \frac{{\left( \alpha  \right)_{m_1  + ... + m_r } \left( {\beta _1 }
\right)_{m_1 }  \cdot  \cdot  \cdot \left( {\beta _r } \right)_{m_r } }}{{\left( {\gamma _1 }
\right)_{m_1 }  \cdot  \cdot  \cdot \left( {\gamma _r } \right)_{m_r } m_1 ! \cdot
\cdot  \cdot m_r !}}x_1^{m_1 }  \cdot  \cdot  \cdot x_r^{m_r } , \\
\,\,\,\,\,\,\,\,\,\,\,\,\,\,\,\,\,\,\,\,\,\,\,\,\,\,\,\,\,\,\,\,\,\,\,\,\,\,\,\,\,\,\,\,\,\,
\,\,\,\,\,\,\,\,\,\,\,\,\,\,\,\,\,\,\,\,\,\,\,\,\,\,\,\,\,\,\,\,\,\,\,\,\,\,\,\,\,\,\,\,\,\,
\,\,\,\,\,\,\,\,\,\,\,\,\,\,\,\,\,\,\,\,\,\,\,\,\,\,\,\,\,\,\,\,\,\,\,\,\,\,\,\,\,
\left( {\left| {x_1 } \right| +  \cdot  \cdot  \cdot  + \left|
{x_r } \right| < 1} \right),
\end{array}\eqno (1.1)
$$
$$
\begin{array}{l}
F_B^{\left( r \right)} \left( {\alpha _1 ,...,\alpha _r ;\beta _1
,...,\beta _r ; \gamma ;x_1 ,...,x_r } \right) = \sum\limits_{m_1
,...,m_r  = 0}^\infty  {} \displaystyle \frac{{\left( {\alpha _1 }
\right)_{m_1 }  \cdot  \cdot  \cdot \left( {\alpha _r }
\right)_{m_r } \left( {\beta _1 } \right)_{m_1 }  \cdot  \cdot
\cdot \left( {\beta _r } \right)_{m_r } }}{{\left( \gamma
\right)_{m_1  + ... + m_r } m_1 ! \cdot  \cdot  \cdot m_r
!}}x_1^{m_1 } \cdot  \cdot  \cdot x_r^{m_r } , \\
\,\,\,\,\,\,\,\,\,\,\,\,\,\,\,\,\,\,\,\,\,\,\,\,\,\,\,\,\,\,\,\,\,\,\,\,\,\,\,\,\,\,\,\,\,\,\,\,\,\,\,\,\,
\,\,\,\,\,\,\,\,\,\,\,\,\,\,\,\,\,\,\,\,\,\,\,\,\,\,\,\,\,\,\,\,\,\,\,\,\,\,\,\,\,\,\,\,\,\,\,\,\,\,\,\,\,
\,\,\,\,\,\,\,\,\,\,\,\,\,\,\,\,\,\,\,\,\,\,\,\,\,\,\left( {\max \left[ {\left| {x_1 } \right|,...,
\,\left| {x_r } \right|} \right] < 1} \right), \\
\end{array} \eqno (1.2)
$$
$$
\begin{array}{l}
F_C^{\left( r \right)} \left( {\alpha ,\beta ;\gamma _1
,...,\gamma _r ;x_1 ,...,x_r } \right) = \sum\limits_{m_1 ,...,m_r
= 0}^\infty  {} \displaystyle \frac{{\left( \alpha  \right)_{m_1
+ ... + m_r } \left( \beta  \right)_{m_1  + ... + m_r } }}{{\left(
{\gamma _1 } \right)_{m_1 }  \cdot  \cdot \cdot \left( {\gamma _r
} \right)_{m_r } m_1 ! \cdot  \cdot  \cdot m_r !}}x_1^{m_1 }
\cdot  \cdot  \cdot x_r^{m_r } , \\
\,\,\,\,\,\,\,\,\,\,\,\,\,\,\,\,\,\,\,\,\,\,\,\,\,\,\,\,\,\,\,\,\,\,\,\,\,\,\,\,\,\,\,\,\,\,\,\,
\,\,\,\,\,\,\,\,\,\,\,\,\,\,\,\,\,\,\,\,\,\,\,\,\,\,\,\,\,\,\,\,\,\,\,\,\,\,\,\,\,\,\,\,\,\,\,\,
\,\,\,\,\,\,\,\,\,\left( {\left| {\sqrt {x_1 } } \right| +  \cdot  \cdot  \cdot  +
\left| {\sqrt {x_r } } \right| < 1} \right), \\
\end{array} \eqno (1.3)
$$
$$
\begin{array}{l}
F_D^{\left( r \right)} \left( {\alpha ;\beta _1 ,...,\beta _r
;\gamma ;x_1 ,...,x_r } \right) = \displaystyle \sum\limits_{m_1
,...,m_r  = 0}^\infty  {} \frac{{\left( \alpha  \right)_{m_1  +
... + m_r } \left( {\beta _1 } \right)_{m_1 }  \cdot  \cdot  \cdot
\left( {\beta _r } \right)_{m_r } }}{{\left(
\gamma  \right)_{m_1  + ... + m_r } m_1 ! \cdot  \cdot  \cdot m_r !}}x_1^{m_1 }  \cdot  \cdot  \cdot x_r^{m_r } , \\
\,\,\,\,\,\,\,\,\,\,\,\,\,\,\,\,\,\,\,\,\,\,\,\,\,\,\,\,\,\,\,\,\,\,\,\,\,\,\,\,\,\,\,\,\,\,\,\,
\,\,\,\,\,\,\,\,\,\,\,\,\,\,\,\,\,\,\,\,\,\,\,\,\,\,\,\,\,\,\,\,\,\,\,\,\,\,\,\,\,\,\,\,\,\,\,\,
\,\,\,\,\,\,\,\,\,\,\,\,\left( {\max \left[ {\left| {x_1 } \right|,...,\,\left| {x_r } \right|} \right] < 1} \right). \\
\end{array} \eqno (1.4)
$$
Here, and in what follows $\left( \alpha  \right)_m  = \Gamma
\left( {\alpha  + m} \right)/\Gamma \left( \alpha  \right)$
denotes the Pochhammer symbol (or the shifted factorial) for all
admissible (real or complex) values of $\alpha$ and $m$.

For various multivariable hypergeometric functions including the
Lauricella multivariable functions $F_A^{\left( r \right)}$,
$F_B^{\left( r \right)}$, $F_C^{\left( r \right)}$ and
$F_D^{\left( r \right)}$, defined by (1.1)-(1.4), Hasanov and
Srivastava [10, 11] presented a number of decompositions formulas
in terms of such simpler hypergeometric functions as the Gauss and
Appell functions. The main object of this sequel to the works of
Hasanov and Srivastava [10, 11] is to show how some rather
elementary techniques based upon certain inverse pairs of symbolic
operators would lead us easily to several decomposition formulas
associated with Lauricella's hypergeometric function $F_A^{\left(
r \right)}$, $F_B^{\left( r \right)}$, $F_C^{\left( r \right)}$
and $F_D^{\left( r \right)}$ in $r$ variables and with other
multiple hypergeometric functions.

Over six decades ago, Burchnall and Chaundy [3, 4] and Chaundy [5]
systematically presented a number of expansion and decomposition
formulas for some double hypergeometric functions in series of
simpler hypergeometric functions. Their method is based upon the
following inverse pairs of symbolic operators:
$$
\nabla _{xy} \left( h \right): = \frac{{\Gamma \left( h
\right)\Gamma \left( {\delta _1  + \delta _2  + h}
\right)}}{{\Gamma \left( {\delta _1  + h} \right)\Gamma \left(
{\delta _2  + h} \right)}} = \sum\limits_{k = 0}^\infty  {}
\frac{{\left( { - \delta _1 } \right)_k \left( { - \delta _2 }
\right)_k }}{{\left( h \right)_k k!}}, \eqno (1.5)
$$
$$
\begin{array}{l}
\Delta _{xy} \left( h \right): = \displaystyle \frac{{\Gamma
\left( {\delta _1  + h} \right) \Gamma \left( {\delta _2  + h}
\right)}}{{\Gamma \left( h \right)\Gamma \left( {\delta _1  +
\delta _2  + h} \right)}} = \sum\limits_{k = 0}^\infty  {}
\frac{{\left( { - \delta _1 } \right)_k \left( { - \delta _2 }
\right)_k }}{{\left( {1 - h - \delta _1  - \delta _2 } \right)_k k!}} \\
\,\,\,\,\,\,\,\,\,\,\,\,\,\,\,\,\,\,\,\,\,\,\, = \displaystyle
\sum\limits_{k = 0}^\infty  {} \frac{{\left( { - 1} \right)^k
\left( h \right)_{2k} \left( { - \delta _1 } \right)_k \left( { -
\delta _2 } \right)_k }}{{\left( {h + k - 1} \right)_k \left( {h +
\delta _1 }
\right)_k \left( {h + \delta _2 } \right)_k k!}}, \\
\end{array} \eqno (1.6)
$$
and
$$
\begin{array}{l}
\nabla _{xy} \left( h \right)\Delta _{xy} \left( g \right): =
\displaystyle \frac{{\Gamma \left( h \right)\Gamma \left( {\delta
_1  + \delta _2  + h} \right)}}{{\Gamma \left( {\delta _1  + h}
\right)\Gamma \left( {\delta _2  + h} \right)}}\frac{{\Gamma
\left( {\delta _1  + g} \right)\Gamma \left( {\delta _2  + g}
\right)}}{{\Gamma \left( g \right)\Gamma \left( {\delta _1  + \delta _2  + g} \right)}} \\
\,\,\,\,\,\,\,\,\,\,\,\,\,\,\,\,\,\,\,\,\,\,\,\,\,\,\,\,\,\,\,\,\,\,\,\,\,\,\,\,\,\,\,
= \displaystyle \sum\limits_{k = 0}^\infty  {} \frac{{\left( {g -
h} \right)_k \left( g \right)_{2k} \left( { - \delta _1 }
\right)_k \left( { - \delta _2 } \right)_k }}{{\left( {g + k - 1}
\right)_k \left( {g + \delta _1 } \right)_k
\left( {g + \delta _2 } \right)_k k!}} \\
\,\,\,\,\,\,\,\,\,\,\,\,\,\,\,\,\,\,\,\,\,\,\,\,\,\,\,\,\,\,\,\,\,\,\,\,\,\,\,\,\,\,\,
= \displaystyle \sum\limits_{k = 0}^\infty  {} \frac{{\left( {h -
g} \right)_k \left( { - \delta _1 } \right)_k \left( { - \delta _2
} \right)_k }}{{\left( h \right)_k \left( {1 - g - \delta _1  -
\delta _2 } \right)_k k!}}\,\,\,\,\,\,\left( {\delta _1 : =
x\frac{\partial }{{\partial x}};\,\,\delta _2 : = y\frac{\partial }{{\partial y}}} \right). \\
\end{array} \eqno (1.7)
$$
We introduce the following multivariable symbolic operators:
$$
H_{x_1 ,...,x_l } \left( {\alpha ,\beta } \right): = \frac{{\Gamma
\left( \beta  \right)\Gamma \left( {\alpha  + \delta _1  +  \cdot
\cdot  \cdot  + \delta _l } \right)}}{{\Gamma \left( \alpha
\right)\Gamma \left( {\beta  + \delta _1  +  \cdot  \cdot  \cdot
+ \delta _l } \right)}} = \sum\limits_{k_1 , \cdot  \cdot  \cdot
,k_l  = 0}^\infty  {} \frac{{\left( {\beta  - \alpha }
\right)_{k_1  +  \cdot  \cdot  \cdot  + k_l } \left( { - \delta _1
} \right)_{k_1 }  \cdot  \cdot  \cdot \left( { - \delta _l }
\right)_{k_l } }}{{\left( \beta  \right)_{k_1  +  \cdot  \cdot
\cdot  + k_l } k_1 ! \cdot  \cdot  \cdot k_l !}} \eqno (1.8)
$$
and
$$
\bar H_{x_1 ,...,x_l } \left( {\alpha ,\beta } \right): =
\frac{{\Gamma \left( \alpha  \right)\Gamma \left( {\beta  + \delta
_1  +  \cdot  \cdot  \cdot  + \delta _l } \right)}}{{\Gamma \left(
\beta  \right)\Gamma \left( {\alpha  + \delta _1  +  \cdot  \cdot
\cdot  + \delta _l } \right)}} = \sum\limits_{k_1 ,...k_l  =
0}^\infty  {} \frac{{\left( {\beta  - \alpha } \right)_{k_1  +
\cdot  \cdot  \cdot  + k_l } \left( { - \delta _1 } \right)_{k_1 }
\cdot  \cdot  \cdot \left( { - \delta _l } \right)_{k_l }
}}{{\left( {1 - \alpha  - \delta _1  -  \cdot  \cdot  \cdot  -
\delta _l } \right)_{k_1  +  \cdot  \cdot  \cdot  + k_l } k_1 !
\cdot  \cdot  \cdot k_l !}}, \eqno (1.9)
$$
$$
\left( {\delta _j : = x_j \frac{\partial }{{\partial x_j }},\,j =
1,...,l;\,\,l \in N : = \left\{ {1,2,3,...} \right\}} \right)
$$
where we have applied known multiple hypergeometric summation
formulas as follows [1, p. 117]:
$$
\begin{array}{l}
F_D^{\left( r \right)} \left( {\alpha ;\beta _1 ,...,\beta _r
;\gamma ;1,...,1} \right) = \displaystyle \frac{{\Gamma \left(
\gamma  \right)\Gamma \left( {\gamma  - \alpha  - \beta _1  -
\cdot \cdot  \cdot  - \beta _r } \right)}}{{\Gamma \left( {\gamma
- \alpha } \right)\Gamma
\left( {\gamma  - \beta _1  -  \cdot  \cdot  \cdot  - \beta _r } \right)}} \\
\left( {{\mathop{\rm Re}\nolimits} \,\left( {\gamma  - \alpha  -
\beta _1  - \cdot  \cdot  \cdot  - \beta _r } \right) >
0,\,\,\gamma  \notin N_0^ -  : =
\left\{ {0, - 1, - 2, - 3,...} \right\}} \right) \\
\end{array}
$$
for the Lauricella function $F_D^{\left( r \right)} $ in $r$
variables, defined by (1.4).

\section{Families of decompositions formulas for Lauricella functions  }

First of all, it is not difficult to derive the following
applications of the (multivariable) symbolic operators defined by
(1.8) and (1.9):
$$
F_A^{\left( r \right)} \left( {\alpha ;\beta _1,...,\beta _r
;\gamma _1 ,...,\gamma _r ;x_1 ,...,x_r } \right) = H_{x_1
,...,x_r } \left( {\alpha ,\varepsilon } \right)F_A^{\left( r
\right)} \left( {\varepsilon ;\beta _1 ,...,\beta _r ,\gamma _1
,...,\gamma _r ;x_1 ,...,x_r } \right), \eqno (2.1)
$$
$$
F_A^{\left( r \right)} \left( {\alpha ;\beta _1 ,...,\beta _r
;\gamma _1 ,...,\gamma _r ;x_1 ,...,x_r } \right) = \bar H_{x_1
,...,x_r } \left( {\varepsilon ,\alpha } \right)F_A^{\left( r
\right)} \left( {\varepsilon ;\beta _1 ,...,\beta _r ,\gamma _1
,...,\gamma _r ;x_1 ,...,x_r } \right), \eqno (2.2)
$$
$$
F_B^{\left( r \right)} \left( {\alpha _1 ,...,\alpha _r ;\beta _1
,...,\beta _r ,\gamma ;x_1 ,...,x_r } \right) = \bar H_{x_1
,...,x_r } \left( {\gamma ,\varepsilon } \right)F_B^{\left( r
\right)} \left( {\alpha _1 ,...,\alpha _r ;\beta _1 ,...,\beta _r
,\varepsilon ;x_1 ,...,x_r } \right), \eqno (2.3)
$$
$$
F_C^{\left( r \right)} \left( {\alpha ,\beta ;\gamma _1
,...,\gamma _r ;x_1 ,...,x_r } \right) = H_{x_1 ,...,x_r } \left(
{\alpha ,\varepsilon } \right)F_C^{\left( r \right)} \left(
{\varepsilon ,\beta ;\gamma _1 ,...,\gamma _r ;x_1 ,...,x_r }
\right), \eqno (2.4)
$$
$$
F_C^{\left( r \right)} \left( {\alpha ,\beta ;\gamma _1
,...,\gamma _r ;x_1 ,...,x_r } \right) = \bar H_{x_1 ,...,x_r }
\left( {\varepsilon ,\alpha } \right)F_C^{\left( r \right)} \left(
{\varepsilon ,\beta ;\gamma _1 ,...,\gamma _r ;x_1 ,...,x_r }
\right), \eqno (2.5)
$$
$$
F_C^{\left( r \right)} \left( {\alpha ,\beta ;\gamma _1
,...,\gamma _r ;x_1 ,...,x_r } \right) = H_{x_1 ,...,x_r } \left(
{\alpha ,\varepsilon _1 } \right)H_{x_1 ,...,x_r } \left( {\beta
,\varepsilon _2 } \right)F_C^{\left( r \right)} \left(
{\varepsilon _1 ,\varepsilon _2 ;\gamma _1 ,...,\gamma _r ;x_1
,...,x_r } \right), \eqno (2.6)
$$
$$
F_C^{\left( r \right)} \left( {\alpha ,\beta ;\gamma _1
,...,\gamma _r ;x_1 ,...,x_r } \right) = \bar H_{x_1 ,...,x_r }
\left( {\varepsilon _1 ,\alpha } \right)\bar H_{x_1 ,...,x_r }
\left( {\varepsilon _2 ,\beta } \right)F_C^{\left( r \right)}
\left( {\varepsilon _1 ,\varepsilon _2 ;\gamma _1 ,...,\gamma _r
;x_1 ,...,x_r } \right), \eqno (2.7)
$$
$$
F_D^{\left( r \right)} \left( {\alpha ;\beta _1 ,...,\beta _r
;\gamma ;x_1 ,...,x_r } \right) = H_{x_1 ,...,x_r } \left( {\alpha
,\varepsilon } \right)F_D^{\left( r \right)} \left( {\varepsilon
;\beta _1 ,...,\beta _r ;\gamma ;x_1 ,...,x_r } \right), \eqno
(2.8)
$$
$$
F_D^{\left( r \right)} \left( {\alpha ;\beta _1 ,...,\beta _r
;\gamma ;x_1 ,...,x_r } \right) = \bar H_{x_1 ,...,x_r } \left(
{\varepsilon ,\alpha } \right)F_D^{\left( r \right)} \left(
{\varepsilon ;\beta _1 ,...,\beta _r ;\gamma ;x_1 ,...,x_r }
\right), \eqno (2.9)
$$
$$
F_D^{\left( r \right)} \left( {\alpha ;\beta _1 ,...,\beta _r
;\gamma ;x_1 ,...,x_r } \right) = H_{x_1 ,...,x_r } \left(
{\varepsilon ,\gamma } \right)F_D^{\left( r \right)} \left(
{\alpha ;\beta _1 ,...,\beta _r ;\varepsilon ;x_1 ,...,x_r }
\right), \eqno (2.10)
$$
$$
F_D^{\left( r \right)} \left( {\alpha ;\beta _1 ,...,\beta _r
;\gamma ;x_1 ,...,x_r } \right) = H_{x_1 ,...,x_r } \left( {\alpha
,\gamma } \right)\left( {1 - x_1 } \right)^{ - \beta _1 }  \cdot
\cdot  \cdot \left( {1 - x_r } \right)^{ - \beta _r } , \eqno
(2.11)
$$
$$
\left( {1 - x_1 } \right)^{ - \beta _1 }  \cdot  \cdot  \cdot
\left( {1 - x_r } \right)^{ - \beta _r }  = \bar H_{x_1 ,...,x_r }
\left( {\alpha ,\gamma } \right)F_D^{\left( r \right)} \left(
{\alpha ;\beta _1 ,...,\beta _r ;\gamma ;x_1 ,...,x_r } \right).
\eqno (2.12)
$$
In the operator identities (2.6) and (2.7), the superposition of
the operators involved can be represented as follows:
$$
\begin{array}{l}
H_{x_1 ,...,x_r } \left( {\alpha ,\varepsilon _1 } \right)H_{x_1 ,...,x_r } \left( {\beta ,\varepsilon _2 } \right) \\
= \displaystyle \sum\limits_{k_1 , \cdot  \cdot  \cdot ,k_r ,l_1 ,
\cdot  \cdot  \cdot ,l_r  = 0}^\infty  {} \frac{{\left(
{\varepsilon _1  - \alpha } \right)_{k_1  +  \cdot  \cdot  \cdot +
k_r } \left( {\varepsilon _2  - \beta } \right)_{l_1  +  \cdot
\cdot  \cdot  + l_r } \left( \beta \right)_{k_1  +  \cdot  \cdot
\cdot  + k_r } \left( { - \delta _1 } \right)_{k_1  + l_1 } \cdot
\cdot  \cdot \left( { - \delta _r } \right)_{k_r  + l_r }
}}{{\left( {\varepsilon _1 } \right)_{k_1  +  \cdot  \cdot  \cdot
+ k_r } \left( {\varepsilon _2 } \right)_{k_1  +  \cdot \cdot
\cdot  + k_r  + l_1  +  \cdot  \cdot  \cdot  + l_r } k_1 ! \cdot
\cdot  \cdot k_r !l_1 !
\cdot  \cdot  \cdot l_r !}} \\
\end{array} \eqno(2.13)
$$
and
$$
\begin{array}{l}
\bar H_{x_1 ,...,x_r } \left( {\varepsilon _1 ,\alpha } \right)\bar H_{x_1 ,...,x_r }
\left( {\varepsilon _2 ,\beta }\right) \\
= \displaystyle \sum\limits_{k_1 ,...k_r ,l_1 ,...l_r  = 0}^\infty
{} \frac{{\left( { - 1} \right)^{k_1  +  \cdot  \cdot  \cdot  +
k_r } \left( {\alpha  - \varepsilon _1 } \right)_{k_1  +  \cdot
\cdot  \cdot  + k_r } \left( {\beta  - \varepsilon _2 }
\right)_{k_1  +  \cdot  \cdot  \cdot  + k_r  + l_1  +  \cdot
\cdot  \cdot  + l_r } \left( \beta  \right)_{k_1  + ... + k_r }
}}{{\left( {\beta  - \varepsilon _2 }
\right)_{k_1  +  \cdot  \cdot  \cdot  + k_r } k_1 ! \cdot  \cdot  \cdot k_r !l_1 ! \cdot  \cdot  \cdot l_r !}} \\
\times \displaystyle \frac{{\left( { - \delta _1 } \right)_{k_1  +
l_1 }  \cdot  \cdot  \cdot \left( { - \delta _r } \right)_{k_r  +
l_r } }}{{\left( {1 - \varepsilon _1  - \delta _1  -  \cdot  \cdot
\cdot  - \delta _r } \right)_{k_1  +  \cdot  \cdot \cdot  + k_r }
\left( {1 - \varepsilon _2  - \delta _1  -  \cdot  \cdot  \cdot  -
\delta _r } \right)_{k_1  +  \cdot  \cdot  \cdot  + k_r  + l_1  +  \cdot  \cdot  \cdot  + l_r } }}. \\
\end{array}\eqno (2.14)
$$
In view of the know Mellin-Barnes contour integral representations
for the Lauricella functions $F_A^{\left( r \right)}$,
$F_B^{\left( r \right)}$, $F_C^{\left( r \right)}$ and
$F_D^{\left( r \right)}$, it is not difficult to give alternative
proofs of the operator identities (2.1) to (2.12) above by using
the Mellin and the inverse Mellin transformations (see, for
example, [14]). The details involved in these alternative
derivations of the operator identities (2.1) to (2.12) are being
omitted here.\\
By virtue of the derivative formulas for the Lauricella functions,
and also of some standard properties of hypergeometric functions,
we find each of the following decompositions formulas for the
Lauricella functions$F_A^{\left( r \right)}$, $F_B^{\left( r
\right)}$, $F_C^{\left( r \right)}$ and $F_D^{\left( r \right)}$ :
$$
\begin{array}{l}
F_A^{\left( r \right)} \left( {\alpha ;\beta _1 ,...,\beta _r
;\gamma _1 ,...,\gamma _r ;x_1 ,...,x_r } \right) = \displaystyle
\sum\limits_{k_1 , \cdot  \cdot  \cdot ,k_r  = 0}^\infty  {}
\frac{{\left( { - 1} \right)^{k_1  +  \cdot  \cdot \cdot  + k_r }
\left( {\varepsilon  - \alpha } \right)_{k_1  + \cdot  \cdot
\cdot  + k_r } \left( {\beta _1 } \right)_{k_1 } \cdot  \cdot
\cdot \left( {\beta _r } \right)_{k_r } }}{{\left( {\gamma
_1}\right)_{k_1 }\cdot  \cdot  \cdot \left( {\gamma _r } \right)_{k_r } k_1 ! \cdot  \cdot  \cdot k_r !}} \\
\times x_1^{k_1 } x_2^{k_2 }  \cdot  \cdot  \cdot x_r^{k_r }
F_A^{\left( r \right)} \left( {\varepsilon  + k_1  +  \cdot  \cdot
\cdot  + k_r ;\beta _1  + k_1 ,...,\beta _r  + k_r ; \gamma _1  +
k_1 ,...,\gamma _r  + k_r ;x_1 ,...,x_r } \right),
\end{array} \eqno (2.15)
$$
$$
\begin{array}{l}
F_A^{\left( r \right)} \left( {\alpha ;\beta _1 ,...,\beta _r
;\gamma _1 ,...,\gamma _r ;x_1 ,...,x_r } \right) = \displaystyle
\sum\limits_{k_1 ,...k_r  = 0}^\infty  {} \frac{{\left( {\alpha  -
\varepsilon } \right)_{k_1  + \cdot \cdot  \cdot  + k_r } \left(
{\beta _1 } \right)_{k_1 } \left( {\beta _2 } \right)_{k_2 }
\cdot \cdot  \cdot \left( {\beta _r } \right)_{k_r } }}{{\left(
{\gamma _1 } \right)_{k_1 } \left( {\gamma _2 }
\right)_{k_2 }  \cdot  \cdot  \cdot \left( {\gamma _r } \right)_{k_r } k_1 ! \cdot  \cdot  \cdot k_r !}} \\
\times x_1^{k_1 }  \cdot  \cdot  \cdot x_r^{k_r } F_A^{\left( r
\right)} \left( {\varepsilon ; \beta _1  + k_1 ,...,\beta _r  +
k_r ;\gamma _1  + k_1 ,...,\gamma _r  + k_r ;x_1 ,...,x_r }
\right),
\end{array} \eqno (2.16)
$$
$$
\begin{array}{l}
F_B^{\left( r \right)} \left( {\alpha _1 ,...,\alpha _r ;\beta _1 ,...,\beta _r ;\gamma ;x_1 ,...,x_r } \right) \\
= \displaystyle \sum\limits_{k_1 , \cdot  \cdot  \cdot ,k_r  =
0}^\infty  {} \frac{{\left( { - 1} \right)^{k_1  +  \cdot  \cdot
\cdot  + k_r } \left( {\gamma  - \varepsilon } \right)_{k_1  +
\cdot  \cdot  \cdot  + k_r } \left( {\alpha _1 } \right)_{k_1 }
\cdot  \cdot  \cdot \left( {\alpha _r } \right)_{k_r } \left(
{\beta _1 } \right)_{k_1 }  \cdot \cdot  \cdot \left( {\beta _r }
\right)_{k_r } }}{{\left( \gamma  \right)_{k_1  +  \cdot  \cdot
\cdot  + k_r } \left( \varepsilon  \right)_{k_1  +  \cdot  \cdot  \cdot  + k_r } k_1 ! \cdot  \cdot  \cdot k_r !}} \\
\times x_1^{k_1 }  \cdot  \cdot  \cdot x_r^{k_r } F_B^{\left( r \right)} \left( {\alpha _1  + k_1 ,...,
\alpha _r  + k_r ;\beta _1  + k_1 ,...,\beta _r  + k_r ;\varepsilon  + k_1  +  \cdot  \cdot
\cdot  + k_r ;x_1 ,...,x_r } \right), \\
\end{array} \eqno (2.17)
$$
$$
\begin{array}{l}
F_C^{\left( r \right)} \left( {\alpha ,\beta ;\gamma _1
,...,\gamma _r ;x_1 ,...,x_r } \right) = \displaystyle
\sum\limits_{k_1 , \cdot  \cdot  \cdot ,k_r  = 0}^\infty  {}
\frac{{\left( { - 1} \right)^{k_1  + \cdot  \cdot \cdot  + k_r }
\left( {\varepsilon  - \alpha } \right)_{k_1  + \cdot  \cdot \cdot
+ k_r } \left( \beta  \right)_{k_1  +  \cdot \cdot  \cdot  + k_r }
}}{{\left( {\gamma _1 } \right)_{k_1 } \cdot  \cdot  \cdot \left(
{\gamma _r } \right)_{k_r } k_1 ! \cdot \cdot
\cdot k_r !}}x_1^{k_1 }  \cdot  \cdot  \cdot x_r^{k_r }  \\
\times F_C^{\left( r \right)} \left( {\varepsilon  + k_1  +  \cdot  \cdot
\cdot  + k_r ,\beta  + k_1  +  \cdot  \cdot  \cdot  + k_r ;\gamma _1  + k_1 ,...,
\gamma _r  + k_r ;x_1 ,...,x_r } \right), \\
\end{array} \eqno (2.18)
$$
$$
\begin{array}{l}
F_C^{\left( r \right)} \left( {\alpha ,\beta ;\gamma _1
,...,\gamma _r ;x_1 ,...,x_r } \right) = \displaystyle
\sum\limits_{k_1 ,...k_r  = 0}^\infty  {} \frac{{\left( {\alpha  -
\varepsilon } \right)_{k_1  + \cdot \cdot  \cdot  + k_r } \left(
\beta  \right)_{k_1  +  \cdot  \cdot \cdot  + k_r } }}{{\left(
{\gamma _1 } \right)_{k_1 }  \cdot \cdot  \cdot \left( {\gamma _r
} \right)_{k_r } k_1 ! \cdot  \cdot
\cdot k_r !}}x_1^{k_1 }  \cdot  \cdot  \cdot x_r^{k_r }  \\
\times F_C^{\left( r \right)} \left( {\varepsilon ,\beta  + k_1  +
\cdot  \cdot  \cdot  + k_r ;\gamma _1  + k_1 ,...,\gamma _r  + k_r ;x_1 ,...,x_r } \right), \\
\end{array}\eqno (2.19)
$$
$$
\begin{array}{l}
F_C^{\left( r \right)} \left( {\alpha ,\beta ;\gamma _1
,...,\gamma _r ;x_1 ,...,x_r } \right) = \displaystyle
\sum\limits_{k_1 , \cdot  \cdot  \cdot ,k_r ,l_1 , \cdot  \cdot
\cdot ,l_r  = 0}^\infty  {} \frac{{\left( { - 1} \right)^{k_1  +
\cdot  \cdot  \cdot  + k_r  + l_1  +  \cdot \cdot  \cdot  + l_r }
\left( {\varepsilon _1  - \alpha } \right)_{k_1  +  \cdot  \cdot
\cdot  + k_r } }}{{\left( {\varepsilon _1 } \right)_{k_1  +  \cdot
\cdot  \cdot  + k_r } \left( {\gamma _1 } \right)_{k_1  + l_1 }
\cdot  \cdot  \cdot \left( {\gamma _r } \right)_{k_r  + l_r } }} \\
\displaystyle \frac{{\left( {\varepsilon _2  - \beta }
\right)_{l_1 +  \cdot  \cdot  \cdot  + l_r } \left( \beta
\right)_{k_1  + \cdot  \cdot  \cdot  + k_r } \left( {\varepsilon
_1 } \right)_{k_1 +  \cdot  \cdot  \cdot  + k_r  + l_1  +  \cdot
\cdot \cdot  + l_r } }}{{k_1 ! \cdot  \cdot  \cdot k_r !l_1 !
\cdot  \cdot  \cdot l_r !}}x_1^{k_1  + l_1 }
\cdot  \cdot  \cdot x_r^{k_r  + l_r }  \\
\displaystyle \times F_C^{\left( r \right)} \left( {\varepsilon _1
+ k_1  +  \cdot  \cdot  \cdot  + k_r  + l_1  + \cdot  \cdot  \cdot
+ l_r ,\varepsilon _2  + k_1  +  \cdot  \cdot  \cdot  + k_r  + l_1
+  \cdot \cdot  \cdot  + l_r ;} \right. \\
\,\,\,\,\,\,\,\,\,\,\,\,\,\,\,\,\,\,\,\,\,\,\,\,\,\,\,\,\,\,\,\,\,\,\,\,\,\,\,\,\,\,\,\,\,\,
\,\,\,\,\,\,\,\,\,\,\,\,\,\,\,\,\,\,\,\,\,\,\,\,\,\,\,\,\,\,\,\,\,\,\,\,
\left. {\gamma _1  + k_1  + l_1 ,...,\gamma _r  + k_r  + l_r ;x_1 ,...,x_r } \right), \\
\end{array} \eqno (2.20)
$$
$$
\begin{array}{l}
F_C^{\left( r \right)} \left( {\alpha ,\beta ;\gamma _1 ,...,\gamma _r ;x_1 ,...,x_r } \right) \\
= \displaystyle \sum\limits_{k_1 ,...k_r ,l_1 ,...l_r  = 0}^\infty
{} \frac{{\left( { - 1} \right)^{l_1  + \cdot  \cdot  \cdot  + l_r
} \left( {\alpha  - \varepsilon _1 } \right)_{k_1  + \cdot  \cdot
\cdot  + k_r } \left( {\beta  - \varepsilon _2 } \right)_{k_1  +
\cdot  \cdot  \cdot  + k_r  + l_1  +  \cdot  \cdot  \cdot  + l_r }
\left( \beta  \right)_{k_1  + ... + k_r } }}{{\left( {\beta  -
\varepsilon _2 }\right)_{k_1  +  \cdot  \cdot  \cdot  + k_r }
\left( {\gamma _1 } \right)_{k_1  + l_1 } \cdot  \cdot  \cdot
\left( {\gamma _r } \right)_{k_r  + l_r } k_1 ! \cdot  \cdot
\cdot k_r !l_1 !\cdot  \cdot  \cdot l_r !}}x_1^{k_1  + l_1 }  \cdot  \cdot  \cdot x_r^{k_r  + l_r }  \\
\times F_C^{\left( r \right)} \left( {\varepsilon _1  + l_1  +  \cdot  \cdot  \cdot  + l_r ,
\varepsilon _2  + l_1  +  \cdot  \cdot  \cdot  + l_r ;\gamma _1  + k_1  + l_1 ,...,
\gamma _r  + k_r  + l_r ;x_1 ,...,x_r } \right), \\
\end{array} \eqno (2.21)
$$
$$
\begin{array}{l}
F_D^{\left( r \right)} \left( {\alpha ;\beta _1 ,...,\beta _r
;\gamma ;x_1 ,...,x_r } \right) = \displaystyle \sum\limits_{k_1 ,
\cdot  \cdot  \cdot ,k_r  = 0}^\infty  {} \frac{{\left( { - 1}
\right)^{k_1  + \cdot  \cdot  \cdot  + k_r } \left( {\varepsilon -
\alpha } \right)_{k_1  +  \cdot  \cdot  \cdot  + k_r } \left(
{\beta _1 } \right)_{k_1 }  \cdot  \cdot  \cdot \left( {\beta _2 }
\right)_{k_2 } }}{{\left( \gamma \right)_{k_1  +  \cdot  \cdot
\cdot  + k_r } k_1 !
\cdot  \cdot  \cdot k_r !}}x_1^{k_1 }\cdot  \cdot  \cdot x_r^{k_r }  \\
\times F_D^{\left( r \right)} \left( {\varepsilon  + k_1  +  \cdot  \cdot
\cdot  + k_r ;\beta _1  + k_1 ,...,\beta _r  + k_r ;\gamma  + k_1  +
\cdot  \cdot  \cdot  + k_r ;x_1 ,...,x_r } \right), \\
\end{array} \eqno (2.22)
$$
$$
\begin{array}{l}
F_D^{\left( r \right)} \left( {\alpha ;\beta _1 ,...,\beta _r
;\gamma ;x_1 ,...,x_r } \right) = \displaystyle \sum\limits_{k_1
,...k_r  = 0}^\infty  {} \frac{{\left( {\alpha  - \varepsilon }
\right)_{k_1  +  \cdot  \cdot  \cdot  + k_r } \left( {\beta _1 }
\right)_{k_1 } \cdot  \cdot  \cdot \left( {\beta _r } \right)_{k_r
} }}{{\left( \gamma \right)_{k_1  +  \cdot  \cdot  \cdot  + k_r }
k_1 ! \cdot  \cdot  \cdot k_r !}}x_1^{k_1 }
\cdot  \cdot  \cdot x_r^{k_r }  \\
\times F_D^{\left( r \right)} \left( {\varepsilon ;\beta _1  + k_1 ,...,\beta _r  + k_r ;
\gamma  + k_1  +  \cdot  \cdot  \cdot  + k_r ;x_1 ,...,x_r } \right), \\
\end{array} \eqno (2.23)
$$
$$
\begin{array}{l}
F_D^{\left( r \right)} \left( {\alpha ;\beta _1 ,...,\beta _r ;\gamma ;x_1 ,...,x_r } \right) \\
= \displaystyle \sum\limits_{k_1 , \cdot  \cdot  \cdot ,k_r  = 0}^\infty  {} \frac{{\left( { - 1}
\right)^{k_1  +  \cdot  \cdot  \cdot  + k_r } \left( {\gamma  - \varepsilon } \right)_{k_1  +
\cdot  \cdot  \cdot  + k_r } \left( \alpha  \right)_{k_1  +  \cdot  \cdot  \cdot  + k_r }
\left( {\beta _1 } \right)_{k_1 }  \cdot  \cdot  \cdot \left( {\beta _r } \right)_{k_r } }}{{\left(
\gamma  \right)_{k_1  +  \cdot  \cdot  \cdot  + k_r } \left( \varepsilon  \right)_{k_1  +
\cdot  \cdot  \cdot  + k_r } k_1 ! \cdot  \cdot  \cdot k_r !}}x_1^{k_1 }  \cdot  \cdot  \cdot x_r^{k_r }  \\
\times F_D^{\left( r \right)} \left( {\alpha  + k_1  +  \cdot  \cdot  \cdot  + k_r ;\beta _1  + k_1 ,...,
\beta _r  + k_r ;\varepsilon  + k_1  +  \cdot  \cdot  \cdot  + k_r ;x_1 ,...,x_r } \right), \\
\end{array} \eqno (2.24)
$$
$$
\begin{array}{l}
F_D^{\left( r \right)} \left( {\alpha ;\beta _1 ,...,\beta _r
;\gamma ;x_1 ,...,x_r } \right)\\
= \left( {1 - x_1 } \right)^{ - \beta _1 }  \cdot  \cdot  \cdot
\left( {1 - x_r } \right)^{ - \beta _r } \displaystyle F_D^{\left(
r \right)} \left( {\gamma  - \alpha ;\beta _1 ,...,\beta _r
;\gamma ;{{x_1 } \over {x_1  - 1}},...,{{x_r } \over {x_r  - 1}}}
\right), \end{array} \eqno (2.25)
$$
$$
\begin{array}{l}
\left( {1 - x_1 } \right)^{ - \beta _1 }  \cdot  \cdot  \cdot
\left( {1 - x_r } \right)^{ - \beta _r }  = \displaystyle
\sum\limits_{k_1 ,...k_r  = 0}^\infty  {} {{\left( {\gamma  -
\alpha } \right)_{k_1 +  \cdot  \cdot  \cdot  + k_r } \left(
{\beta _1 } \right)_{k_1 } \cdot  \cdot  \cdot \left( {\beta _r }
\right)_{k_r } } \over {\left( \gamma  \right)_{k_1  + \cdot
\cdot  \cdot  + k_r } k_1 ! \cdot  \cdot  \cdot k_r !}}x_1^{k_1 }
\cdot  \cdot  \cdot x_r^{k_r }   \\
\displaystyle \times F_D^{\left( r \right)} \left( {\alpha ;\beta
_1  + k_1 ,...,\beta _r  + k_r ;\gamma  + k_1  +  \cdot \cdot
\cdot  + k_r ;x_1 ,...,x_r } \right),
\end{array} \eqno (2.26)
$$
Our operational derivations of the decomposition formulas (2.15)
to (2.26) would indeed run parallel to those presented in the
earlier works, which we have already cited in the preceding
sections. In addition to the various operator expressions, we also
make use of the following operator identities [18, p. 93]:
$$
\left( {\delta  + \alpha } \right)_n \left\{ {f\left( \xi \right)}
\right\} = \xi ^{1 - \alpha } {{d^n } \over {d\xi ^n }}\left\{
{\xi ^{\alpha  + n - 1} f\left( \xi  \right)} \right\} \eqno
(2.27)
$$
$$
\left( {\delta : = \xi {d \over {d\xi }};\,\,\alpha  \in ;\,\,n
\in _0 : =  \cup \left\{ 0 \right\};\,\,: = \left\{ {1,2,3,...}
\right\}} \right)
$$
and
$$
\left( { - \delta } \right)_n \left\{ {f\left( \xi  \right)}
\right\} = \left( { - 1} \right)^n \xi ^n {{d^n } \over {d\xi ^n
}}\left\{ {f\left( \xi  \right)} \right\},\,\,\left( {\delta : =
\xi {d \over {d\xi }};\,\,n \in _0 } \right) \eqno (2.28)
$$
for every analytic function $f\left( \xi  \right)$. Many other
analogous decomposition formulas can similarly be derived for the
Lauricella functions $F_A^{\left( r \right)}$, $F_B^{\left( r
\right)}$, $F_C^{\left( r \right)}$ and $F_D^{\left( r \right)}$
in $r$ variables, but with various different parametric
constraints.

\section{Basic operator identities for Lauricella functions in the three-variable cases  }

In this section, we consider the Lauricella functions $F_A^{\left(
3 \right)}$, $F_B^{\left( 3 \right)}$, $F_C^{\left( 3 \right)}$
and $F_D^{\left( 3 \right)}$ in three variables. We reiterate the
aforementioned fact that numerous other analogous decompositions
formulas can similarly be derived for the Lauricella triple
hypergeometric functions $F_A^{\left( 3 \right)}$, $F_B^{\left( 3
\right)}$, $F_C^{\left( 3 \right)}$ and $F_D^{\left( 3 \right)}$
with different parametric constrains. For example, each of the
following operational representations would lead us a
decomposition formula for a special $F_A^{\left( 3 \right)}$,
$F_B^{\left( 3 \right)}$, $F_C^{\left( 3 \right)}$ and
$F_D^{\left( 3 \right)}$ in three variables.
$$
F_A \left( {\alpha ;\beta _1 ,\beta _2 ,\beta _3 ;\gamma _1
,\gamma _2 ,\gamma _3 ;x,y,z} \right) = H_x \left( {\beta _1
,\varepsilon _1 } \right)F_A \left( {\alpha ;\varepsilon _1 ,\beta
_2 ,\beta _3 ;\gamma _1 ,\gamma _2 ,\gamma _3 ;x,y,z} \right),
\eqno (3.1)
$$
$$
F_A \left( {\alpha ;\beta _1 ,\beta _2 ,\beta _3 ;\gamma _1
,\gamma _2 ,\gamma _3 ;x,y,z} \right) = \bar H_x \left(
{\varepsilon _1 ,\beta _1 } \right)F_A \left( {\alpha ;\varepsilon
_1 ,\beta _2 ,\beta _3 ;\gamma _1 ,\gamma _2 ,\gamma _3 ;x,y,z}
\right), \eqno (3.2)
$$
$$
F_A \left( {\alpha ;\beta _1 ,\beta _2 ,\beta _3 ;\gamma _1
,\gamma _2 ,\gamma _3 ;x,y,z} \right) = H_x \left( {\beta _1
,\varepsilon _1 } \right)H_y \left( {\beta _2 ,\varepsilon _2 }
\right)F_A \left( {\alpha ;\varepsilon _1 ,\varepsilon _2 ,\beta
_3 ;\gamma _1 ,\gamma _2 ,\gamma _3 ;x,y,z} \right), \eqno (3.3)
$$
$$
F_A \left( {\alpha ;\beta _1 ,\beta _2 ,\beta _3 ;\gamma _1
,\gamma _2 ,\gamma _3 ;x,y,z} \right) = \bar H_x \left(
{\varepsilon _1 ,\beta _1 } \right)\bar H_y \left( {\varepsilon _2
,\beta _2 } \right)F_A \left( {\alpha ;\varepsilon _1 ,\varepsilon
_2 ,\beta _3 ;\gamma _1 ,\gamma _2 ,\gamma _3 ;x,y,z} \right),
\eqno (3.4)
$$
$$
F_A \left( {\alpha ;\beta _1 ,\beta _2 ,\beta _3 ;\gamma _1
,\gamma _2 ,\gamma _3 ;x,y,z} \right)  \\ = H_x \left( {\beta _1
,\varepsilon _1 } \right)H_y \left( {\beta _2 ,\varepsilon _2 }
\right)H_z \left( {\beta _3 ,\varepsilon _3 } \right)F_A \left(
{\alpha ;\varepsilon _1 ,\varepsilon _2 ,\varepsilon _3 ;\gamma _1
,\gamma _2 ,\gamma _3 ;x,y,z} \right), \eqno (3.5)
$$
$$
F_A \left( {\alpha ;\beta _1 ,\beta _2 ,\beta _3 ;\gamma _1
,\gamma _2 ,\gamma _3 ;x,y,z} \right)  \\ = \bar H_x \left(
{\varepsilon _1 ,\beta _1 } \right)\bar H_y \left( {\varepsilon _2
,\beta _2 } \right)\bar H_z \left( {\varepsilon _3 ,\beta _3 }
\right)F_A \left( {\alpha ;\varepsilon _1 ,\varepsilon _2
,\varepsilon _3 ;\gamma _1 ,\gamma _2 ,\gamma _3 ;x,y,z} \right),
\eqno (3.6)
$$
$$
F_A \left( {\alpha ;\beta _1 ,\beta _2 ,\beta _3 ;\gamma _1
,\gamma _2 ,\gamma _3 ;x,y,z} \right) = H_x \left( {\beta _1
,\gamma _1 } \right)\left( {1 - x} \right)^{ - \alpha } F_2 \left(
{\alpha ;\beta _2 ,\beta _3 ;\gamma _2 ,\gamma _3 ;{y \over {1 -
x}},{z \over {1 - x}}} \right), \eqno (3.7)
$$
$$
\left( {1 - x} \right)^{ - \alpha } F_2 \left( {\alpha ;\beta _2
,\beta _3 ;\gamma _2 ,\gamma _3 ;{y \over {1 - x}},{z \over {1 -
x}}} \right) = \bar H_x \left( {\beta _1 ,\gamma _1 } \right)F_A
\left( {\alpha ;\beta _1 ,\beta _2 ,\beta _3 ;\gamma _1 ,\gamma _2
,\gamma _3 ;x,y,z} \right), \eqno (3.8)
$$
$$
F_A \left( {\alpha ;\beta _1 ,\beta _2 ,\beta _3 ;\gamma _1
,\gamma _2 ,\gamma _3 ;x,y,z} \right) = H_x \left( {\beta _1
,\gamma _1 } \right)H_y \left( {\beta _2 ,\gamma _2 }
\right)\left( {1 - x - y} \right)^{ - \alpha } F\left( {\alpha
,\beta _3 ;\gamma _3 ;{z \over {1 - x - y}}} \right), \eqno (3.9)
$$
$$
\left( {1 - x - y} \right)^{ - \alpha } F\left( {\alpha ,\beta _3
;\gamma _3 ;{z \over {1 - x - y}}} \right) = \bar H_x \left(
{\beta _1 ,\gamma _1 } \right)\bar H_y \left( {\beta _2 ,\gamma _2
} \right)F_A \left( {\alpha ;\beta _1 ,\beta _2 ,\beta _3 ;\gamma
_1 ,\gamma _2 ,\gamma _3 ;x,y,z} \right),\eqno (3.10)
$$
$$
F_A \left( {\alpha ;\beta _1 ,\beta _2 ,\beta _3 ;\gamma _1
,\gamma _2 ,\gamma _3 ;x,y,z} \right) = H_x \left( {\beta _1
,\gamma _1 } \right)H_y \left( {\beta _2 ,\gamma _2 } \right)H_z
\left( {\beta _3 ,\gamma _3 } \right)\left( {1 - x - y - z}
\right)^{ - \alpha } ,\eqno (3.11)
$$
$$
\left( {1 - x - y - z} \right)^{ - \alpha }  = \bar H_x \left(
{\beta _1 ,\gamma _1 } \right)\bar H_y \left( {\beta _2 ,\gamma _2
} \right)\bar H_z \left( {\beta _3 ,\gamma _3 } \right)F_A \left(
{\alpha ;\beta _1 ,\beta _2 ,\beta _3 ;\gamma _1 ,\gamma _2
,\gamma _3 ;x,y,z} \right),\eqno (3.12)
$$
$$
F_B \left( {\alpha _1 ,\alpha _2 ,\alpha _3 ;\beta _1 ,\beta _2
,\beta _3 ;\gamma ;x,y,z} \right) = H_x \left( {\alpha _1
,\varepsilon _1 } \right)F_B \left( {\varepsilon _1 ,\alpha _2
,\alpha _3 ;\beta _1 ,\beta _2 ,\beta _3 ;\gamma ;x,y,z}
\right),\eqno (3.13)
$$
$$
F_B \left( {\alpha _1 ,\alpha _2 ,\alpha _3 ;\beta _1 ,\beta _2
,\beta _3 ;\gamma ;x,y,z} \right) = \bar H_x \left( {\varepsilon
_1 ,\alpha _1 } \right)F_B \left( {\varepsilon _1 ,\alpha _2
,\alpha _3 ;\beta _1 ,\beta _2 ,\beta _3 ;\gamma ;x,y,z}
\right),\eqno (3.14)
$$
$$
F_B \left( {\alpha _1 ,\alpha _2 ,\alpha _3 ;\beta _1 ,\beta _2
,\beta _3 ;\gamma ;x,y,z} \right) = H_x \left( {\alpha _1
,\varepsilon _1 } \right)H_y \left( {\alpha _2 ,\varepsilon _2 }
\right)F_B \left( {\varepsilon _1 ,\varepsilon _2 ,\alpha _3
;\beta _1 ,\beta _2 ,\beta _3 ;\gamma ;x,y,z} \right),\eqno (3.15)
$$
$$
F_B \left( {\alpha _1 ,\alpha _2 ,\alpha _3 ;\beta _1 ,\beta _2
,\beta _3 ;\gamma ;x,y,z} \right) = \bar H_x \left( {\varepsilon
_1 ,\alpha _1 } \right)\bar H_y \left( {\varepsilon _2 ,\alpha _2
} \right)F_B \left( {\varepsilon _1 ,\varepsilon _2 ,\alpha _3
;\beta _1 ,\beta _2 ,\beta _3 ;\gamma ;x,y,z} \right),\eqno (3.16)
$$
$$
\begin{array}{l}
F_B \left( {\alpha _1 ,\alpha _2 ,\alpha _3 ;\beta _1 ,\beta _2 ,\beta _3 ;\gamma ;x,y,z} \right) \\
\,\,\,\,\,\,\,\,\,\,\,\,\,\,\,\,\,\,\,\,\,\,\,\,\,\,\,\,\,\,\,\,\,\,\,\,\,\,\,\,\,\,\,\,\,\,\,\,\,\,\,\,\,\,\,\,\,\,\,\,
= H_x \left( {\alpha _1 ,\varepsilon _1 } \right)H_y \left(
{\alpha _2 ,\varepsilon _2 } \right)H_z \left( {\alpha _3
,\varepsilon _3 } \right)F_B \left( {\varepsilon _1 ,\varepsilon
_2 ,\varepsilon _3 ;
\beta _1 ,\beta _2 ,\beta _3 ;\gamma ;x,y,z} \right), \\
\end{array}\eqno (3.17)
$$
$$
\begin{array}{l}
F_B \left( {\alpha _1 ,\alpha _2 ,\alpha _3 ;\beta _1 ,\beta _2 ,\beta _3 ;\gamma ;x,y,z} \right) \\
\,\,\,\,\,\,\,\,\,\,\,\,\,\,\,\,\,\,\,\,\,\,\,\,\,\,\,\,\,\,\,\,\,\,\,\,\,\,\,\,\,\,\,\,\,\,\,\,\,\,\,\,\,\,\,\,\,\,=
\bar H_x \left( {\varepsilon _1 ,\alpha _1 } \right)\bar H_y
\left( {\varepsilon _2 ,\alpha _2 } \right)\bar H_z \left(
{\varepsilon _3 ,\alpha _3 } \right)F_B \left( {\varepsilon _1
,\varepsilon _2 ,
\varepsilon _3 ;\beta _1 ,\beta _2 ,\beta _3 ;\gamma ;x,y,z} \right), \\
\end{array}  \eqno (3.18)
$$
$$
F_C \left( {\alpha ,\beta ;\gamma _1 ,\gamma _2 ,\gamma _3 ;x,y,z}
\right) = H_x \left( {\varepsilon _1 ,\gamma _1 } \right)F_C
\left( {\alpha ,\beta ;\varepsilon _1 ,\gamma _2 ,\gamma _3
;x,y,z} \right),\eqno (3.19)
$$
$$
F_C \left( {\alpha ,\beta ;\gamma _1 ,\gamma _2 ,\gamma _3 ;x,y,z}
\right) = H_x \left( {\varepsilon _1 ,\gamma _1 } \right)H_y
\left( {\varepsilon _2 ,\gamma _2 } \right)F_C \left( {\alpha
,\beta ;\varepsilon _1 ,\varepsilon _2 ,\gamma _3 ;x,y,z}
\right),\eqno (3.20)
$$
$$
F_C \left( {\alpha ,\beta ;\gamma _1 ,\gamma _2 ,\gamma _3 ;x,y,z}
\right) = H_x \left( {\varepsilon _1 ,\gamma _1 } \right)H_y
\left( {\varepsilon _2 ,\gamma _2 } \right)H_z \left( {\varepsilon
_3 ,\gamma _3 } \right)F_C \left( {\alpha ,\beta ;\varepsilon _1
,\varepsilon _2 ,\varepsilon _3 ;x,y,z} \right),\eqno (3.21)
$$
$$
F_D \left( {\alpha ;\beta _1 ,\beta _2 ,\beta _3 ;\gamma ;x,y,z}
\right) = H_x \left( {\beta _1 ,\varepsilon _1 } \right)F_D \left(
{\alpha ;\varepsilon _1 ,\beta _2 ,\beta _3 ;\gamma ;x,y,z}
\right),\eqno (3.22)
$$
$$
F_D \left( {\alpha ;\beta _1 ,\beta _2 ,\beta _3 ;\gamma ;x,y,z}
\right) = \bar H_x \left( {\varepsilon _1 ,\beta _1 } \right)F_D
\left( {\alpha ;\varepsilon _1 ,\beta _2 ,\beta _3 ;\gamma ;x,y,z}
\right),\eqno (3.23)
$$
$$
F_D \left( {\alpha ;\beta _1 ,\beta _2 ,\beta _3 ;\gamma ;x,y,z}
\right) = H_x \left( {\beta _1 ,\varepsilon _1 } \right)H_y \left(
{\beta _2 ,\varepsilon _2 } \right)F_D \left( {\alpha ;\varepsilon
_1 ,\varepsilon _2 ,\beta _3 ;\gamma ;x,y,z} \right),\eqno (3.24)
$$
$$
F_D \left( {\alpha ;\beta _1 ,\beta _2 ,\beta _3 ;\gamma ;x,y,z}
\right) = \bar H_x \left( {\varepsilon _1 ,\beta _1 } \right)\bar
H_y \left( {\varepsilon _2 ,\beta _2 } \right)F_D \left( {\alpha
;\varepsilon _1 ,\varepsilon _2 ,\beta _3 ;\gamma ;x,y,z}
\right),\eqno (3.25)
$$
$$
F_D \left( {\alpha ;\beta _1 ,\beta _2 ,\beta _3 ;\gamma ;x,y,z}
\right) = H_x \left( {\beta _1 ,\varepsilon _1 } \right)H_y \left(
{\beta _2 ,\varepsilon _2 } \right)H_z \left( {\beta _3
,\varepsilon _3 } \right)F_D \left( {\alpha ;\varepsilon _1
,\varepsilon _2 ,\varepsilon _3 ;\gamma ;x,y,z} \right),\eqno
(3.26)
$$
$$
F_D \left( {\alpha ;\beta _1 ,\beta _2 ,\beta _3 ;\gamma ;x,y,z}
\right) = \bar H_x \left( {\varepsilon _1 ,\beta _1 } \right)\bar
H_y \left( {\varepsilon _2 ,\beta _2 } \right)\bar H_z \left(
{\varepsilon _3 ,\beta _3 } \right)F_D \left( {\alpha ;\varepsilon
_1 ,\varepsilon _2 ,\varepsilon _3 ;\gamma ;x,y,z} \right),\eqno
(3.27)
$$
where, for notational convenience
$$
F_A : = F_A^{\left( 3 \right)}, F_B : = F_B^{\left( 3 \right)},
F_C : = F_C^{\left( 3 \right)}, F_C : = F_C^{\left( 3 \right)},
F_2 : = F_A^{\left( 2 \right)}. \eqno (3.28)
$$
By virtue of the derivative formulas for the Lauricella functions,
and also of some standard properties of hypergeometric functions,
we find each of the following decompositions formulas for the
Lauricella functions $F_A^{\left( 3 \right)}$, $F_B^{\left( 3
\right)}$, $F_C^{\left( 3 \right)}$ and $F_D^{\left( 3 \right)}$:
$$
\begin{array}{l}
F_A \left( {\alpha ;\beta _1 ,\beta _2 ,\beta _3 ;\gamma _1 ,\gamma _2 ,\gamma _3 ;x,y,z} \right) \\
\,\,\,\,\,\,\,\,\,\,\,\,\,\,\,\,\,\,\,\,\,\,\,\,\,\,\,\,\,\,\,\,\,\,\,\,\,\,\,\,\,\,\,\,\,\,\,\,\,\,
= \displaystyle \sum\limits_{i = 0}^\infty {} \frac{{\left( { - 1}
\right)^i \left( {\varepsilon _1 - \beta _1 } \right)_i \left(
\alpha \right)_i }}{{\left( {\gamma _1 } \right)_i i!}}x^i F_A
\left( {\alpha  + i;\varepsilon _1  + i,\beta _2 ,\beta _3 ;\gamma _1  + i,\gamma _2 ,\gamma _3 ;x,y,z} \right), \\
\end{array} \eqno (3.29)
$$
$$
\begin{array}{l}
F_A \left( {\alpha ;\beta _1 ,\beta _2 ,\beta _3 ;\gamma _1
,\gamma _2 ,\gamma _3 ;x,y,z} \right) = \displaystyle
\sum\limits_{i = 0}^\infty {} \frac{{\left( {\beta _1  -
\varepsilon _1 } \right)_i \left( \alpha \right)_i }}{{\left(
{\gamma _1 } \right)_i i!}}x^i F_A \left( {\alpha  + i;\varepsilon
_1 ,\beta _2 ,\beta _3 ;\gamma _1  + i,\gamma _2 ,\gamma _3 ;x,y,z} \right), \\
\end{array} \eqno (3.30)
$$
$$
\begin{array}{l}
F_A \left( {\alpha ;\beta _1 ,\beta _2 ,\beta _3 ;\gamma _1
,\gamma _2 ,\gamma _3 ;x,y,z} \right) = \displaystyle
\sum\limits_{i,j = 0}^\infty  {} \frac{{\left( { - 1} \right)^{i +
j} \left( \alpha  \right)_{i + j} \left( {\varepsilon _1  - \beta
_1 } \right)_i \left( {\varepsilon _2  - \beta _2 } \right)_j
}}{{\left( {\gamma _1 } \right)_i \left( {\gamma _2 }\right)_j i!j!}}x^i y^j  \\
\,\,\,\,\,\,\,\,\,\,\,\,\,\,\,\,\,\,\,\,\,\,\,\,\,\,\,\,\,\,\,\,\,\,\,\,\,\,\,\,\,\,\,\,\,\,\,\,\,\,\,\,\,\,\,\,\,
\,\,\,\,\,\,\,\,\,\,\,\,\,\,\,\,\,\,\,\,\,\,\,\,\,\,\,\,\,\,\,\,\,
\times F_A \left( {\alpha + i + j;\varepsilon _1 + i,\varepsilon
_2  + j,\beta _3 ;\gamma _1 + i, \gamma _2  + j,\gamma _3 ;x,y,z} \right), \\
\end{array}\eqno (3.31)
$$
$$
\begin{array}{l}
F_A \left( {\alpha ;\beta _1 ,\beta _2 ,\beta _3 ;\gamma _1
,\gamma _2 ,\gamma _3 ;x,y,z} \right) = \displaystyle
\sum\limits_{i,j = 0}^\infty  {} {{\left( \alpha \right)_{i + j}
\left( {\beta _1  - \varepsilon _1 } \right)_i \left( {\beta _2  -
\varepsilon _2 } \right)_j } \over {\left( {\gamma _1 } \right)_i
\left( {\gamma _2 } \right)_j i!j!}}x^i y^j  \\
\,\,\,\,\,\,\,\,\,\,\,\,\,\,\,\,\,\,\,\,\,\,\,\,\,\,\,\,\,\,\,\,\,\,\,\,\,\,\,\,\,\,\,\,\,\,\,\,\,\,\,\,\,\,\,\,\,
\,\,\,\,\,\,\,\,\,\,\,\,\,\,\,\,\,\,\,\,\,\,\,\,\,\,\,\,\,\,\,\,
\times F_A \left( {\alpha  + i + j;\varepsilon _1 ,\varepsilon _2
,\beta _3 ;\gamma _1  + i,\gamma _2  + j,\gamma _3 ;x,y,z}
\right),\end{array} \eqno (3.32)
$$
$$
\begin{array}{l}
F_A \left( {\alpha ;\beta _1 ,\beta _2 ,\beta _3 ;\gamma _1
,\gamma _2 ,\gamma _3 ;x,y,z} \right)  \\
\,\,\,\,\,\,\,\,\,\,\,\,\,\,\,\,\,\,\,\,\,\,\,\,\,\,\,\,\,\,\,\,\,\,\,\,\,\,\,\,\,\,\,\,\,\,\,\,\,\,\,\,\,\,\,\,\,
= \displaystyle \sum\limits_{i,j,k = 0}^\infty  {} {{\left( { - 1}
\right)^{i + j + k} \left( \alpha \right)_{i + j + k} \left(
{\varepsilon _1 - \beta _1 } \right)_i \left( {\varepsilon _2  -
\beta _2 } \right)_j \left( {\varepsilon _3  - \beta _3 }
\right)_k } \over {\left( {\gamma _1 } \right)_i \left( {\gamma _2
} \right)_j \left( {\gamma _3 } \right)_k i!j!k!}}x^i y^j z^k   \\
\,\,\,\,\,\,\,\,\,\,\,\,\,\,\,\,\,\,\,\,\,\,\,\,\,\,\,\,\,\,\,\,\,\,\,\,\,\,\,\,\,\,\,\,\,\,\,\,\,\,\,\,\,\,\,\,\,
\times F_A \left( {\alpha  + i + j + k;\varepsilon _1  +
i,\varepsilon _2 + j,\varepsilon _3  + k;\gamma _1  + i,\gamma _2
+ j,\gamma _3  + k;x,y,z} \right),\end{array} \eqno (3.33)
$$
$$
\begin{array}{l}
F_A \left( {\alpha ;\beta _1 ,\beta _2 ,\beta _3 ;\gamma _1
,\gamma _2 ,\gamma _3 ;x,y,z} \right) = \displaystyle
\sum\limits_{i,j,k = 0}^\infty  {} \frac{{\left( \alpha
\right)_{i + j + k} \left( {\beta _1  - \varepsilon _1 } \right)_i
\left( {\beta _2  - \varepsilon _2 } \right)_j \left( {\beta _3  -
\varepsilon _3 } \right)_k }}{{\left( {\gamma _1 } \right)_i
\left( {\gamma _2 }\right)_j \left( {\gamma _3 } \right)_k i!j!k!}}x^i y^j z^k  \\
\,\,\,\,\,\,\,\,\,\,\,\,\,\,\,\,\,\,\,\,\,\,\,\,\,\,\,\,\,\,\,\,\,\,\,\,\,\,\,\,\,\,\,\,\,\,\,\,\,\,\,\,\,\,\,\,\,
\,\,\,\,\,\,\,\,\,\,\,\,\,\,\,\,\,\,\,\,\,\,\,\,\,\,\,\,\,\,\,\,
\times F_A \left( {\alpha  + i + j + k;\varepsilon _1 ,\varepsilon
_2 ,\varepsilon _3 ;\gamma _1  + i,\gamma _2  + j,\gamma _3  + k;x,y,z} \right), \\
\end{array} \eqno (3.34)
$$
$$
\begin{array}{l}
F_A \left( {\alpha ;\beta _1 ,\beta _2 ,\beta _3 ;\gamma _1
,\gamma _2 ,\gamma _3 ;x,y,z} \right)  \\
\,\,\,\,\,\,\,\,\,\,\,\,\,\,\,\,\,\,\,\,\,\,\,\,\,\,\,\,\,\,\,\,
=\displaystyle \left( {1 - x} \right)^{ - \alpha } \sum\limits_{i
= 0}^\infty {} {{\left( \alpha \right)_i \left( {\gamma _1  -
\beta _1 } \right)_i } \over {\left( {\gamma _1 } \right)_i
i!}}\left( {{x \over {x - 1}}} \right)^i F_2 \left( {\alpha  +
i;\beta _2 ,\beta _3 ;\gamma _2 ,\gamma _3 ;{y \over {1 - x}},{z
\over {1 - x}}} \right),\end{array} \eqno (3.35)
$$
$$
\begin{array}{l}
\displaystyle \left( {1 - x} \right)^{ - \alpha } F_2 \left(
{\alpha ;\beta _2 ,\beta _3 ;\gamma _2 ,\gamma _3 ;{y \over {1 -
x}},{z \over {1 - x}}} \right)  \\
\,\,\,\,\,\,\,\,\,\,\,\,\,\,\,\,\,\,\,\,\,\,\,\,\,\,\,\,\,\,\,\,
\,\,\,\,\,\,\,\,\,\,\,\,\,\,\,\,\,\,\,\,\,\,\,\,\,\,\,\,\,\,\,\, =
\displaystyle \sum\limits_{i = 0}^\infty  {} {{\left( \alpha
\right)_i \left( {\gamma _1  - \beta _1 } \right)_i } \over
{\left( {\gamma _1 } \right)_i i!}}x^i F_A \left( {\alpha  +
i;\beta _1 ,\beta _2 ,\beta _3 ;\gamma _1  + i,\gamma _2 ,\gamma
_3 ;x,y,z} \right),\end{array} \eqno (3.36)
$$
$$
\begin{array}{l}
F_A \left( {\alpha ;\beta _1 ,\beta _2 ,\beta _3 ;\gamma _1
,\gamma _2 ,\gamma _3 ;x,y,z} \right) = \left( {1 - x - y}
\right)^{ - \alpha } \displaystyle \sum\limits_{i,j = 0}^\infty {}
\frac{{\left( { - 1} \right)^{i + j} \left( \alpha \right)_{i + j}
\left( {\gamma _1  - \beta _1 } \right)_i \left( {\gamma _2  -
\beta _2 } \right)_j }}{{\left( {\gamma _1 } \right)_i \left({\gamma _2 } \right)_j i!j!}}\\
\,\,\,\,\,\,\,\,\,\,\,\,\,\,\,\,\,\,\,\,\,\,\,\,\,\,\,\,\,\,\,\,
\,\,\,\,\,\,\,\,\,\,\,\,\,\,\,\,\,\,\,\,\,\,\,\,\,\,\,\,\,\,\,\,
\,\,\,\,\,\,\,\,\,\,\, \displaystyle \times \left( {\frac{x}{{1 -
x - y}}} \right)^i \left( {\frac{y}{{1 - x - y}}} \right)^j
F\left( {\alpha + i + j,\beta _3 ;\gamma _3 ; \frac{z}{{1 - x - y}}} \right), \\
\end{array} \eqno (3.37)
$$
$$
\begin{array}{l}
\displaystyle \left( {1 - x - y} \right)^{ - \alpha } F\left( {\alpha ,\beta _3 ;\gamma _3 ;
\frac{z}{{1 - x - y}}} \right) \\
\,\,\,\,\,\,\,\,\,\,\,\,\, = \displaystyle \sum\limits_{i,j =
0}^\infty  {} \frac{{\left( \alpha \right)_{i + j} \left( {\gamma
_1  - \beta _1 } \right)_i \left( {\gamma _2  - \beta _2 }
\right)_j }}{{\left( {\gamma _1 } \right)_i \left( {\gamma _2 }
\right)_j i!j!}}x^i y^j F_A \left( {\alpha  + i + j;\beta _1
,\beta _2 ,\beta _3 ;\gamma _1  +
i,\gamma _2  + j, \gamma _3 ;x,y,z} \right), \\
\end{array} \eqno (3.38)
$$
$$
\begin{array}{l}
\displaystyle F_A \left( {\alpha ;\beta _1 ,\beta _2 ,\beta _3
;\gamma _1 ,\gamma _2 ,\gamma _3 ;x,y,z} \right) = \left( {1 - x -
y - z} \right)^{ - \alpha }   \\ \,\,\,\,\,\, \times \displaystyle
F_A \left( {\alpha ;\gamma _1 - \beta _1 ,\gamma _2 - \beta _2
,\gamma _3  - \beta _3 ;\gamma _1 ,\gamma _2 ,\gamma _3 ;{x \over
{x + y + z - 1}},{y \over {x + y + z - 1}},{z \over {x + y + z -
1}}} \right),\end{array} \eqno (3.39)
$$
$$
\begin{array}{l}
\displaystyle \left( {1 - x - y - z} \right)^{ - \alpha } =
\displaystyle \sum\limits_{i,j,k = 0}^\infty  {} {{\left( \alpha
\right)_{i + j + k} \left( {\gamma _1  - \beta _1 } \right)_i
\left( {\gamma _2 - \beta _2 } \right)_j \left( {\gamma _3  -
\beta _3 } \right)_k } \over {\left( {\gamma _1 } \right)_i \left(
{\gamma _2 } \right)_j \left( {\gamma _3 } \right)_k i!j!k!}}x^i y^j z^k   \\
\,\,\,\,\,\,\,\,\,\,\,\,\,\,\,\,\,\,\,\,\,\,\,\,\,\,\,\,\,\,\,\,
\,\,\,\,\,\,\,\,\,\,\,\,\,\,\,\,\, \times F_A \left( {\alpha  + i
+ j + k;\beta _1 ,\beta _2 ,\beta _3 ;\gamma _1  + i,\gamma _2  +
j,\gamma _3  + k;x,y,z} \right),\end{array} \eqno (3.40)
$$
$$
\begin{array}{l}
F_B \left( {\alpha _1 ,\alpha _2 ,\alpha _3 ;\beta _1 ,\beta _2 ,\beta _3 ;\gamma ;x,y,z} \right) \\
\,\,\,\,\,\,\,\,\,\,\,\,\,\,\,\,\,\,\,\,\,\,\,\,\,\,\,\,\,\,\,\,
\,\,\,\,\,\,\,\,\,\,\,\,\,\,\,\,\,= \displaystyle \sum\limits_{i =
0}^\infty  {} \frac{{\left( { - 1} \right)^i \left( {\varepsilon
_1  - \alpha _1 } \right)_i \left( {\beta _1 } \right)_i
}}{{\left( \gamma  \right)_i i!}}x^i F_B \left( {\varepsilon _1  +
i, \alpha _2 ,\alpha _3 ;\beta _1  + i,\beta _2 ,\beta _3 ;\gamma  + i;x,y,z} \right), \\
\end{array} \eqno (3.41)
$$
$$
\begin{array}{l}
F_B \left( {\alpha _1 ,\alpha _2 ,\alpha _3 ;\beta _1 ,\beta _2 ,\beta _3 ;\gamma ;x,y,z} \right) \\
\,\,\,\,\,\,\,\,\,\,\,\,\,\,\,\,\,\,\,\,\,\,\,\,\,\,\,\,\,\,\,\,
\,\,\,\,\,\,\,\,\,\,\,\,\,\,\,\,\,\,\,\,\,\,\,\,\,\,\,\,\,\,\,\, =
\displaystyle \sum\limits_{i = 0}^\infty  {} \frac{{\left( {\alpha
_1  - \varepsilon _1 } \right)_i \left( {\beta _1 } \right)_i
}}{{\left( {\gamma _1 } \right)_i i!}}x^i F_B \left( {\varepsilon
_1 ,\alpha _2 ,\alpha _3 ;
\beta _1  + i,\beta _2 ,\beta _3 ;\gamma  + i;x,y,z} \right), \\
\end{array} \eqno (3.42)
$$
$$
\begin{array}{l}
F_B \left( {\alpha _1 ,\alpha _2 ,\alpha _3 ;\beta _1 ,\beta _2
,\beta _3 ;\gamma ;x,y,z} \right) = \displaystyle \sum\limits_{i,j
= 0}^\infty  {} \frac{{\left( { - 1} \right)^{i + j} \left(
{\varepsilon _1  - \alpha _1 } \right)_i \left( {\varepsilon _2  -
\alpha _2 } \right)_j \left( {\beta _1 } \right)_i \left( {\beta
_2 } \right)_j }}{{\left( \gamma  \right)_{i + j} i!j!}}x^i y^j  \\
\,\,\,\,\,\,\,\,\,\,\,\,\,\,\,\,\,\,\,\,\,\,\,\,\,\,\,\,\,\,\,\,
\,\,\,\,\,\,\,\,\,\,\,\,\,\,\,\,\,\,\,\,\,\,\,\,\,\,\,\,\,\,\,\,
\,\,\,\,\,\,\,\,\,\,\,\,\,\,\,\,\, \times F_B \left( {\varepsilon
_1  + i,\varepsilon _2  + j,\alpha _3 ;\beta _1  + i,\beta _2  +
j,\beta _3 ; \gamma  + i + j;x,y,z} \right), \\
\end{array} \eqno (3.43)
$$
$$
\begin{array}{l}
F_B \left( {\alpha _1 ,\alpha _2 ,\alpha _3 ;\beta _1 ,\beta _2 ,\beta _3 ;\gamma ;x,y,z} \right) \\
= \displaystyle \sum\limits_{i,j = 0}^\infty  {} \frac{{\left(
{\alpha _1  - \varepsilon _1 } \right)_i \left( {\alpha _2  -
\varepsilon _2 } \right)_i \left( {\beta _1 } \right)_i \left(
{\beta _2 } \right)_j }}{{\left( {\gamma _1 } \right)_{i + j}
i!j!}}x^i y^j F_B \left( {\varepsilon _1 ,\varepsilon _2 ,\alpha
_3 ;\beta _1  + i,\beta _2  + j,
\beta _3 ;\gamma  + i + j;x,y,z} \right), \\
\end{array} \eqno (3.44)
$$
$$
\begin{array}{l}
F_B \left( {\alpha _1 ,\alpha _2 ,\alpha _3 ;\beta _1 ,\beta _2 ,\beta _3 ;\gamma ;x,y,z} \right) \\
\,\,\,\,\,\,\,\,\,\,\,\,\,\,\,\,\,\,\,\,\,\,\,\,\,\,\,\,\,\,\,\, =
\displaystyle \sum\limits_{i,j,k = 0}^\infty  {} \frac{{\left( { -
1} \right)^{i + j + k} \left( {\varepsilon _1  - \alpha _1 }
\right)_i \left( {\varepsilon _2  - \alpha _2 } \right)_j \left(
{\varepsilon _3  - \alpha _3 } \right)_k \left( {\beta _1 }
\right)_i \left( {\beta _2 } \right)_j \left( {\beta _3 }
\right)_k }}{{\left( \gamma \right)_{i + j + k} i!j!k!}}x^i y^j z^k  \\
\,\,\,\,\,\,\,\,\,\,\,\,\,\,\,\,\,\,\,\,\,\,\,\,\,\,\,\,\,\,\,\,
\displaystyle \times F_B \left( {\varepsilon _1  + i,\varepsilon
_2 + j,\varepsilon _3  + k;\beta _1  + i,\beta _2  + j,
\beta _3  + k;\gamma  + i + j + k;x,y,z} \right), \\
\end{array} \eqno (3.45)
$$
$$
\begin{array}{l}
F_B \left( {\alpha _1 ,\alpha _2 ,\alpha _3 ;\beta _1 ,\beta _2
,\beta _3 ;\gamma ;x,y,z} \right) \\
\,\,\,\,\,\,\,\,\,\,\,\,\,\,\,\,\,\,\,\,\,\,\,\,\,\,\,\,\,\,\,\,
\,\,\,\,\,\,\,\,\,\,\,\,\,\,\,\,\,\,\,\,\,\,\,\,\,\,\,\,\,\,\,\, =
\displaystyle \sum\limits_{i,j,k = 0}^\infty  {\frac{{\left(
{\alpha _1  - \varepsilon _1 } \right)_i \left( {\alpha _2  -
\varepsilon _2 } \right)_i \left( {\alpha _3  - \varepsilon _3 }
\right)_k \left( {\beta _1 } \right)_i \left( {\beta _2 }
\right)_j \left( {\beta _3 } \right)_k }}{{\left( {\gamma _1 }
\right)_{i + j + k} i!j!k!}}x^i y^j z^k }  \\
\,\,\,\,\,\,\,\,\,\,\,\,\,\,\,\,\,\,\,\,\,\,\,\,\,\,\,\,\,\,\,\,
\,\,\,\,\,\,\,\,\,\,\,\,\,\,\,\,\,\,\,\,\,\,\,\,\,\,\,\,\,\,\,\,
\displaystyle \times F_B \left( {\varepsilon _1 ,\varepsilon _2
,\varepsilon _3 ;\beta _1  + i,\beta _2  + j,\beta _3  + k;
\gamma  + i + j + k;x,y,z} \right), \\
\end{array} \eqno (3.46)
$$
$$
\begin{array}{l}
F_C \left( {\alpha ,\beta ;\gamma _1 ,\gamma _2 ,\gamma _3 ;x,y,z}
\right) = \displaystyle \sum\limits_{i = 0}^\infty  {}
\frac{{\left( { - 1} \right)^i \left( \alpha  \right)_i \left(
\beta  \right)_i \left( {\gamma _1  - \varepsilon _1 } \right)_i
}}{{\left( {\gamma _1 } \right)_i \left( {\varepsilon _1 }
\right)_i i!}}x^i F_C \left( {\alpha  + i,\beta  + i;\varepsilon _1  + i,\gamma _2 ,\gamma _3 ;x,y,z} \right), \\
\end{array} \eqno (3.47)
$$
$$
\begin{array}{l}
F_C \left( {\alpha ,\beta ;\gamma _1 ,\gamma _2 ,\gamma _3 ;x,y,z}
\right) = \displaystyle \sum\limits_{i,j = 0}^\infty  {}
\frac{{\left( { - 1} \right)^{i + j} \left( \alpha  \right)_{i +
j} \left( \beta  \right)_{i + j} \left( {\gamma _1  - \varepsilon
_1 } \right)_i \left( {\gamma _2  - \varepsilon _2 } \right)_j
}}{{\left( {\gamma _1 } \right)_i
\left( {\gamma _2 } \right)_j \left( {\varepsilon _1 } \right)_i \left( {\varepsilon _2 } \right)_j i!j!}}x^i y^j  \\
\,\,\,\,\,\,\,\,\,\,\,\,\,\,\,\,\,\,\,\,\,\,\,\,\,\,\,\,\,\,\,\,
\,\,\,\,\,\,\,\,\,\,\,\,\,\,\,\,\,\,\,\,\,\,\,\,\,\,\,\,\,\,\,\,\,\,\,\,\,
\times F_C \left( {\alpha  + i + j,\beta  + i + j;\varepsilon _1  + i,\varepsilon _2  + j,\gamma _3 ;x,y,z} \right), \\
\end{array} \eqno (3.48)
$$
$$
\begin{array}{l}
F_C \left( {\alpha ,\beta ;\gamma _1 ,\gamma _2 ,\gamma _3 ;x,y,z} \right) \\
\,\,\,\,\,\,\,\,\,\,\,\,\,\,\,\,\,\,\,\,\,\,\,\,\,\,\,\,\,\,\,\, =
\displaystyle \sum\limits_{i,j,k = 0}^\infty  {} \frac{{\left( { -
1} \right)^{i + j + k} \left( \alpha  \right)_{i + j + k} \left(
\beta  \right)_{i + j + k} \left( {\gamma _1  - \varepsilon _1 }
\right)_i \left( {\gamma _2  - \varepsilon _2 } \right)_j \left(
{\gamma _3  - \varepsilon _3 } \right)_k }}{{\left( {\gamma _1 }
\right)_i \left( {\gamma _2 } \right)_j \left( {\gamma _3 }
\right)_k \left( {\varepsilon _1 } \right)_i \left( {\varepsilon
_2 }\right)_j \left( {\varepsilon _3 } \right)_k i!j!k!}}x^i y^j z^k  \\
\,\,\,\,\,\,\,\,\,\,\,\,\,\,\,\,\,\,\,\,\,\,\,\,\,\,\,\,\,\,\,\,
\times F_C \left( {\alpha  + i + j + k,\beta  + i + j +
k;\varepsilon _1  + i,\varepsilon _2  + j,
\varepsilon _3  + k;x,y,z} \right), \\
\end{array} \eqno (3.49)
$$
$$
F_D \left( {\alpha ;\beta _1 ,\beta _2 ,\beta _3 ;\gamma ;x,y,z}
\right) = \displaystyle \sum\limits_{i = 0}^\infty  {}
\frac{{\left( { - 1} \right)^i \left( \alpha  \right)_i \left(
{\varepsilon _1  - \beta _1 } \right)_i }}{{\left( \gamma
\right)_i i!}}x^i F_D \left( {\alpha  + i;\varepsilon _1  +
i,\beta _2 ,\beta _3 ;\gamma  + i;x,y,z} \right),\eqno (3.50)
$$
$$
F_D \left( {\alpha ;\beta _1 ,\beta _2 ,\beta _3 ;\gamma ;x,y,z}
\right) = \displaystyle \sum\limits_{i = 0}^\infty  {}
\frac{{\left( \alpha \right)_i \left( {\beta _1  - \varepsilon _1
} \right)_i }}{{\left( \gamma  \right)_i i!}}x^i F_D \left(
{\alpha  + i;\varepsilon _1 ,\beta _2 ,\beta _3 ;\gamma  +
i;x,y,z} \right),\eqno (3.51)
$$
$$
\begin{array}{l}
F_D \left( {\alpha ;\beta _1 ,\beta _2 ,\beta _3 ;\gamma ;x,y,z} \right) \\
= \displaystyle \sum\limits_{i,j = 0}^\infty  {} \frac{{\left( { -
1} \right)^{i + j} \left( \alpha  \right)_{i + j} \left(
{\varepsilon _1  - \beta _1 } \right)_i \left( {\varepsilon _2  -
\beta _2 } \right)_j }}{{\left( \gamma \right)_{i + j} i!j!}}x^i
y^j F_D \left( {\alpha  + i + j;\varepsilon _1  + i,\varepsilon _2
+ j,\beta _3 ;\gamma  + i + j;x,y,z} \right),
\end{array} \eqno (3.52)
$$
$$
\begin{array}{l}
F_D \left( {\alpha ;\beta _1 ,\beta _2 ,\beta _3 ;\gamma ;x,y,z} \right) \\
\,\,\,\,\,\,\,\,\,\,\,\,\,\,\,\,\,\,\,\,\,\,\,\,\,\,\,\,\,\,\,\,
=\displaystyle \sum\limits_{i,j = 0}^\infty  {} \frac{{\left(
\alpha  \right)_{i + j} \left( {\beta _1  - \varepsilon _1 }
\right)_i \left( {\beta _2  - \varepsilon _2 } \right)_j
}}{{\left( \gamma  \right)_{i + j} i!j!}}x^i y^j F_D
\left( {\alpha  + i + j;\varepsilon _1 ,\varepsilon _2 ,\beta _3 ;\gamma  + i + j;x,y,z} \right), \\
\end{array} \eqno (3.53)
$$
$$
\begin{array}{l}
F_D \left( {\alpha ;\beta _1 ,\beta _2 ,\beta _3 ;\gamma ;x,y,z}
\right) =\displaystyle \sum\limits_{i,j,k = 0}^\infty  {}
\frac{{\left( { - 1} \right)^{i + j + k} \left( \alpha  \right)_{i
+ j + k} \left( {\varepsilon _1  - \beta _1 } \right)_i \left(
{\varepsilon _2  - \beta _2 } \right)_j \left( {\varepsilon _3  -
\beta _3 }\right)_k }}{{\left( \gamma  \right)_{i + j + k} i!j!k!}}x^i y^j z^k  \\
\,\,\,\,\,\,\,\,\,\,\,\,\,\,\,\,\,\,\,\,\,\,\,\,\,\,\,\,\,\,\,\,
\,\,\,\,\,\,\,\,\,\,\,\,\,\,\,\,\,\,\,\,\,\,\,\,\,\,\,\,\,\,\,\,\,\,\,\,\,\,
\times F_D \left( {\alpha  + i + j + k;\varepsilon _1  +
i,\varepsilon _2  + j,\varepsilon _3  + k;
\gamma  + i + j + k;x,y,z} \right), \\
\end{array} \eqno (3.54)
$$
$$
\begin{array}{l}
F_D \left( {\alpha ;\beta _1 ,\beta _2 ,\beta _3 ;\gamma ;x,y,z}
\right) = \displaystyle \sum\limits_{i,j,k = 0}^\infty  {}
\frac{{\left( \alpha  \right)_{i + j + k} \left( {\beta _1  -
\varepsilon _1 } \right)_i \left( {\beta _2  - \varepsilon _2 }
\right)_j \left( {\beta _3  - \varepsilon _3 } \right)_k
}}{{\left( \gamma \right)_{i + j + k} i!j!k!}}x^i y^j z^k  \\
\,\,\,\,\,\,\,\,\,\,\,\,\,\,\,\,\,\,\,\,\,\,\,\,\,\,\,\,\,\,\,\,
\,\,\,\,\,\,\,\,\,\,\,\,\,\,\,\,\,\,\,\,\,\,\,\,\,\,\,\,\,\,\,\,\,\,\,\,\,\,
\times F_D \left( {\alpha  + i + j + k;\varepsilon _1 ,\varepsilon
_2 ,\varepsilon _3 ;\gamma  + i + j + k;x,y,z}
\right). \\
\end{array} \eqno (3.55)
$$

\section{Demonstrations of some the decompositions formulas for Lauricella functions in the three-variable cases }

The various decomposition formulas for the Lauricella functions
$F_A $, $F_B $, $F_C $ and $F_D $ in three variables (which are
presented here and in other places in the previously cited
literature) can be proven fairly simply by suitably applying
superposition of the inverse pairs of symbolic operators
introduced in Section 1. As an example, we shall briefly indicate
the proof of the decomposition formula (3.54). For the
three-variable Lauricella function $F_D$, it is not difficult to
show from (3.26) that
$$
\begin{array}{l}
F_D \left( {\alpha ;\beta _1 ,\beta _2 ,\beta _3 ;\gamma ;x,y,z} \right) \\
= \displaystyle \sum\limits_{i = 0}^\infty  {} \frac{{\left(
{\varepsilon _1  - \beta _1 } \right)_i \left( { - \delta _1 }
\right)_i }}{{\left( {\varepsilon _1 } \right)_i
i!}}\sum\limits_{j = 0}^\infty  {} \frac{{\left( {\varepsilon _2 -
\beta _2 } \right)_j \left( { - \delta _2 } \right)_j }}{{\left(
{\varepsilon _2 } \right)_j j!}}\sum\limits_{k = 0}^\infty  {}
\frac{{\left( {\varepsilon _3  - \beta _3 } \right)_k \left( { -
\delta _3 } \right)_k }}{{\left( {\varepsilon _3 } \right)_k
k!}}F_D \left( {\alpha ;\varepsilon _1 ,\varepsilon _2 ,\varepsilon _3 ;\gamma ;x,y,z} \right), \\
\end{array} \eqno (4.1)
$$
$$
\left( {\delta _1 : = x\frac{\partial }{{\partial x}};\,\,\delta
_2 : = y\frac{\partial }{{\partial y}};\,\,\delta _3 : =
z\frac{\partial }{{\partial z}}} \right).
$$
Furthermore, by a straightforward computation, we have
$$
\displaystyle \left( { - \delta _1 } \right)_i F_D \left( {\alpha
;\varepsilon _1 ,\varepsilon _2 ,\varepsilon _3 ;\gamma ;x,y,z}
\right) = \left( { - 1} \right)^i x^i \frac{{\left( \alpha
\right)_i \left( {\varepsilon _1 } \right)_i }}{{\left( \gamma
\right)_i }}F_D \left( {\alpha  + i;\varepsilon _1  +
i,\varepsilon _2 ,\varepsilon _3 ;\gamma  + i;x,y,z} \right),
\eqno (4.2)
$$

$$
\begin{array}{l}
\left( { - \delta _2 } \right)_j \left( { - \delta _1 } \right)_i F_D \left( {\alpha ;\varepsilon _1 ,
\varepsilon _2 ,\varepsilon _3 ;\gamma ;x,y,z} \right) \\
= \displaystyle \left( { - 1} \right)^{i + j} x^i y^j
\frac{{\left( \alpha  \right)_{i + j} \left( {\varepsilon _1 }
\right)_i \left( {\varepsilon _2 } \right)_j }}{{\left( \gamma
\right)_{i + j} }}F_D \left( {\alpha  + i + j;
\varepsilon _1  + i,\varepsilon _2  + j,\varepsilon _3 ;\gamma  + i + j;x,y,z} \right) \\
\end{array} \eqno (4.3)
$$
and
$$
\begin{array}{l}
\left( { - \delta _3 } \right)_k \left( { - \delta _2 } \right)_j \left( { - \delta _1 } \right)_i F_D
\left( {\alpha ;\varepsilon _1 ,\varepsilon _2 ,\varepsilon _3 ;\gamma ;x,y,z} \right) = \left( { - 1}
\right)^{i + j + k} x^i y^j z^k  \\
\displaystyle \times \frac{{\left( \alpha  \right)_{i + j + k}
\left( {\varepsilon _1 } \right)_i \left( {\varepsilon _2 }
\right)_j \left( {\varepsilon _3 } \right)_k }}{{\left( \gamma
\right)_{i + j + k} }}F_D \left( {\alpha  + i + j + k;
\varepsilon _1  + i,\varepsilon _2  + j,\varepsilon _3  + k;\gamma  + i + j + k;x,y,z} \right). \\
\end{array} \eqno (4.4)
$$
Upon substituting from (4.4) into (4.1), we finally arrive at the
decomposition formula (3.54).

\section{Integral representations via decomposition formulas}

Here in this section, we observe that several integral
representations of the Eulerian type can be deduced also from the
corresponding decomposition formulas of Section 3. For example,
using integral representation ([1], p. 115, (5))
$$
\begin{array}{l}
F_A^{\left( r \right)} \left( {\alpha ;\beta _1 ,\beta _2 ,\beta _3 ,\gamma _1 ,\gamma _2 ,\gamma _3 ;x,y,z} \right) =
\displaystyle \frac{{\Gamma \left( {\gamma _1 } \right)\Gamma \left( {\gamma _2 } \right)\Gamma \left( {\gamma _3 }
\right)}}{{\Gamma \left( {\beta _1 } \right)\Gamma \left( {\beta _2 } \right)\Gamma \left( {\beta _3 } \right)\Gamma
\left( {\gamma _1  - \beta _1 } \right)\Gamma \left( {\gamma _2  - \beta _2 } \right)\Gamma
\left( {\gamma _3  - \beta _3 } \right)}} \\
\int\limits_0^1 {...\int\limits_0^1 {} t_1^{\beta _1  - 1} t_2^{\beta _2  - 1} t_3^{\beta _3  - 1}
\left( {1 - t_1 } \right)^{\gamma _1  - \beta _1  - 1} \left( {1 - t_2 } \right)^{\gamma _2  -
\beta _2  - 1} \left( {1 - t_3 } \right)^{\gamma _3  - \beta _3  - 1} \left( {1 - xt_1  - yt_1  - zt_3 }
\right)^{ - \alpha } dt_1 dt_2 dt_3 } , \\
{\mathop{\rm Re}\nolimits} \,\,\gamma _l  > {\mathop{\rm Re}\nolimits} \,\beta _l \, > 0,\,\,l = 1,2,3 \\
\end{array}\eqno (5.1)
$$
we define
$$
\begin{array}{l}
F_A \left( {\alpha ;\beta _1 ,\beta _2 ,\beta _3 ;\gamma _1 ,\gamma _2 ,\gamma _3 ;x,y,z} \right) =
\displaystyle \frac{{\Gamma \left( {\gamma _1 } \right)\Gamma \left( {\gamma _2 } \right)\Gamma \left( {\gamma _3 }
\right)}}{{\Gamma \left( {\varepsilon _1 } \right)\Gamma \left( {\beta _2 } \right)\Gamma \left( {\beta _3 }
\right)\Gamma \left( {\gamma _1  - \varepsilon _1 } \right)\Gamma \left( {\gamma _2  - \beta _2 } \right)\Gamma
\left( {\gamma _3  - \beta _3 } \right)}} \\
\displaystyle \int\limits_0^1 {...\int\limits_0^1 {}
t_1^{\varepsilon _1  - 1} t_2^{\beta _2  - 1} t_3^{\beta _3  - 1}
\left( {1 - t_1 } \right)^{\gamma _1  - \varepsilon _1  - 1}
\left( {1 - t_2 } \right)^{\gamma _2  - \beta _2  - 1}
\left( {1 - t_3 } \right)^{\gamma _3  - \beta _3  - 1} }  \\
\displaystyle \times \left( {1 - yt_2  - zt_3 } \right)^{ - \alpha
} F\left( {\alpha ,\beta _1 ;\varepsilon _1 ;
\frac{{xt_1 }}{{1 - yt_2  - zt_3 }}} \right)dt_1 dt_2 dt_3 ,\,\,\, \\
{\mathop{\rm Re}\nolimits} \,\,\gamma _1  > {\mathop{\rm Re}\nolimits} \,\,\varepsilon _1  > 0,\,\,{\mathop{\rm Re}
\nolimits} \,\,\gamma _l  > {\mathop{\rm Re}\nolimits} \,\,\beta _l  > 0,\,\,l = 2,3 \\
\end{array}\eqno (5.2)
$$
$$
\begin{array}{l}
F_A \left( {\alpha ;\beta _1 ,\beta _2 ,\beta _3 ;\gamma _1
,\gamma _2 ,\gamma _3 ;x,y,z} \right) = \displaystyle
\frac{{\Gamma \left( {\gamma _1 } \right)\Gamma \left( {\gamma _2
} \right)\Gamma \left( {\gamma _3 } \right)}}{{\Gamma \left(
{\varepsilon _1 } \right)\Gamma \left( {\varepsilon _2 }
\right)\Gamma \left( {\beta _3 } \right)\Gamma \left( {\gamma _1 -
\varepsilon _1 } \right)\Gamma \left( {\gamma _2  -
\varepsilon _2 } \right)\Gamma \left( {\gamma _3  - \beta _3 } \right)}} \\
\displaystyle \times \int\limits_0^1 {...\int\limits_0^1 {}
t_1^{\varepsilon _1  - 1} t_2^{\varepsilon _2  - 1} t_3^{\beta _3
- 1} \left( {1 - t_1 } \right)^{\gamma _1  - \varepsilon _1  - 1}
\left( {1 - t_2 } \right)^{\gamma _2  - \varepsilon _2  - 1}
\left( {1 - t_3 } \right)^{\gamma _3  -
\beta _3  - 1} \left( {1 - xt_1  - yt_1  - zt_3 } \right)^{ - \alpha } }  \\
\displaystyle \times F_2 \left( {\alpha ;\beta _1  - \varepsilon
_1 ,\beta _2  - \varepsilon _2 ;\gamma _1  - \varepsilon _1
,\gamma _2  - \varepsilon _2 ;\frac{{x\left( {1 - t_1 }
\right)}}{{1 - xt_1  - yt_1  - zt_3 }},
\frac{{y\left( {1 - t_2 } \right)}}{{1 - xt_1  - yt_1  - zt_3 }}} \right)dt_1 dt_2 dt_3  \\
\displaystyle {\mathop{\rm Re}\nolimits} \,\,\gamma _l  >
{\mathop{\rm Re}\nolimits} \,\,\varepsilon _l  > 0,
\,\,l = 1,2,\,\,\,{\mathop{\rm Re}\nolimits} \,\,\gamma _3  > {\mathop{\rm Re}\nolimits} \,\beta _3 \, > 0, \\
\end{array}\eqno (5.3)
$$
$$
\begin{array}{l}
F_A \left( {\alpha ;\beta _1 ,\beta _2 ,\beta _3 ;\gamma _1 ,\gamma _2 ,\gamma _3 ;x,y,z} \right) =
\displaystyle \frac{{\Gamma \left( {\gamma _1 } \right)\Gamma \left( {\gamma _2 } \right)\Gamma
\left( {\gamma _3 } \right)}}{{\Gamma \left( {\varepsilon _1 } \right)\Gamma \left( {\varepsilon _2 } \right)\Gamma
\left( {\varepsilon _3 } \right)\Gamma \left( {\gamma _1  - \varepsilon _1 } \right)\Gamma \left( {\gamma _2  -
\varepsilon _2 } \right)\Gamma \left( {\gamma _3  - \varepsilon _3 } \right)}} \\
\displaystyle \times \int\limits_0^1 {...\int\limits_0^1 {}
t_1^{\varepsilon _1 - 1} t_2^{\varepsilon _2  - 1} t_3^{
\varepsilon _3  - 1}\left( {1 - t_1 } \right)^{\gamma _1  -
\varepsilon _1  - 1} \left( {1 - t_2 } \right)^{\gamma _2
-\varepsilon _2  - 1} \left( {1 - t_3 } \right)^{\gamma _3  -
\varepsilon _3  - 1}
\left( {1 - xt_1  - yt_1  - zt_3 } \right)^{ - \alpha } }  \\
\displaystyle \times F_A \left( {\alpha ;\beta _1  - \varepsilon
_1 ,\beta _2  - \varepsilon _2 ,\beta _3  - \varepsilon _3 ;
\gamma _1  - \varepsilon _1 ,\gamma _2  - \varepsilon _2 ,\gamma _3  - \varepsilon _3 ;} \right. \\
\left.
{\,\,\,\,\,\,\,\,\,\,\,\,\,\,\,\,\,\,\,\,\,\,\,\,\,\,\,\,\,\,\,\,\,\,\,\,\,\,\,\,\,\,\,\,\,\,\,\,
\displaystyle \frac{{x\left( {1 - t_1 } \right)}}{{1 - xt_1  -
yt_1 - zt_3 }},\displaystyle \frac{{y\left( {1 - t_2 }
\right)}}{{1 - xt_1  - yt_1  - zt_3 }},\displaystyle
\frac{{z\left( {1 - t_3 } \right)}}{{1 - xt_1  -
yt_1  - zt_3 }}} \right)dt_1 dt_2 dt_3 , \\
{\mathop{\rm Re}\nolimits} \,\,\gamma _l  > {\mathop{\rm Re}\nolimits} \,\,\varepsilon _l  > 0,\,\,l = 1,2,3.\, \\
\end{array} \eqno (5.4)
$$
\textbf{Remark.} Introduced operators can be used for other
hypergeometric functions also.

\end{document}